\documentclass[11pt]{article}
\usepackage{amsmath,amssymb,amsthm,amsxtra,overpic,bbm,bm,epsfig,subfigure}
\usepackage{color}

\usepackage{lscape}

\def\thefootnote{\fnsymbol{footnote}}

\begin{document}

\vspace{0.2cm}

\begin{center}
{\Large\bf A Further Study of the Frampton-Glashow-Yanagida Model for \\ Neutrino Masses, Flavor Mixing and Baryon Number Asymmetry}
\end{center}

\vspace{0.2cm}

\begin{center}
{\bf Jue Zhang $^{a}$} \footnote{E-mail: zhangjue@ihep.ac.cn}
\quad {\bf Shun Zhou $^{a,~b}$} \footnote{E-mail: zhoush@ihep.ac.cn}
\\
{$^a$Institute of High Energy Physics, Chinese Academy of
Sciences, Beijing 100049, China \\
$^b$Center for High Energy Physics, Peking University, Beijing 100871, China}
\end{center}

\vspace{1.5cm}

\begin{abstract}
In light of the latest neutrino oscillation data, we revisit the minimal scenario of type-I seesaw model, in which only two heavy right-handed Majorana neutrinos are introduced to account for both tiny neutrino masses and the baryon number asymmetry in our Universe. In this framework, we carry out a systematic study of the Frampton-Glashow-Yanagida ansatz by taking into account the renormalization-group running of neutrino mixing parameters and the flavor effects in leptogenesis. We demonstrate that the normal neutrino mass ordering is disfavored even in the minimal supersymmetric standard model with a large value of $\tan \beta$, for which the running effects could be significant. Furthermore, it is pointed out that the original scenario with a hierarchical mass spectrum of heavy Majorana neutrinos contradicts with the upper bound derived from a naturalness criterion, and the resonant mechanism with nearly-degenerate heavy Majorana neutrinos can be a possible way out.
\end{abstract}

\begin{flushleft}
\hspace{0.8cm} PACS number(s): 14.60.Pq, 14.60.St, 11.30.Fs
\end{flushleft}

\def\thefootnote{\arabic{footnote}}
\setcounter{footnote}{0}

\newpage

\section{Introduction}

Neutrino oscillation experiments in the past two decades have revealed that neutrinos are actually massive particles and lepton flavors are significantly mixed~\cite{Agashe:2014kda}. In order to accommodate tiny neutrino masses, one can go beyond the minimal Standard Model (SM) and introduce three right-handed neutrinos $N^{}_{i{\rm R}}$ (for $i = 1, 2, 3$), which are singlets under the ${\rm SU(2)}^{}_{\rm L}\times {\rm U(1)}^{}_{\rm Y}$ gauge group of the SM. The most general gauge-invariant Lagrangian relevant for lepton masses and flavor mixing can be written as
\begin{equation}
-{\cal L}^{}_{\rm m} = \overline{\ell^{}_{\rm L}} Y^{}_l H E^{}_{\rm R} + \overline{\ell^{}_{\rm L}} Y^{}_\nu \tilde{H} N^{}_{\rm R} + \frac{1}{2} \overline{N^{\rm c}_{\rm R}} M^{}_{\rm R} N^{}_{\rm R} + {\rm h.c.} \; ,
\label{eq: massterm}
\end{equation}
where $\ell^{}_{\rm L}$ and $\tilde{H} \equiv i\sigma^{}_2 H^*$ denote  the left-handed lepton and Higgs doublets, respectively, while $E^{}_{\rm R}$ the right-handed charged-lepton singlets. In addition, $Y^{}_l$ and $Y^{}_\nu$ stand respectively for the Yukawa coupling matrices of charged leptons and neutrinos, and $M^{}_{\rm R}$ is the Majorana mass matrix for right-handed neutrino singlets. After the Higgs field acquires its vacuum expectation value $\langle H \rangle = v \approx 174~{\rm GeV}$ and the gauge symmetry is spontaneously broken down, the charged-lepton mass matrix is given by $M^{}_l = Y^{}_l v$, while the Dirac neutrino mass matrix is $M^{}_{\rm D} = Y^{}_\nu v$. Since the Majorana mass term for right-handed neutrino singlets is not subject to the electroweak gauge symmetry breaking, the absolute scale of $M^{}_{\rm R}$ could be much higher than the electroweak energy scale $\Lambda^{}_{\rm EW} \sim 100~{\rm GeV}$. Therefore, in the low-energy effective theory with heavy Majorana neutrinos integrated out, the mass matrix of three light neutrinos is given by the famous seesaw formula $M^{}_\nu \approx - M^{}_{\rm D} M^{-1}_{\rm R} M^{\rm T}_{\rm D}$. Given ${\cal O}(M^{}_{\rm D}) \sim \Lambda^{}_{\rm EW}$, one can obtain neutrino masses at the sub-eV level if ${\cal O}(M^{}_{\rm R}) \sim 10^{14}~{\rm GeV}$ is close to the scale of grand unified theories $\Lambda^{}_{\rm GUT} \sim 10^{16}~{\rm GeV}$. In this canonical seesaw model~\cite{Minkowski:1977sc, Yanagida:1979ss, Gell-Mann:1979ss, Glashow:1979ss, Mohapatra:1979ia}, the lightness of ordinary neutrinos can be ascribed to the heaviness of right-handed Majorana neutrinos. Moreover, the mismatch between the diagonalization of $M^{}_l$ and $M^{}_\nu$ leads to lepton flavor mixing.

In the basis where both the charged-lepton mass matrix $M^{}_l = {\rm diag}\{m^{}_e, m^{}_\mu, m^{}_\tau\}$ and the mass matrix of heavy Majorana neutrinos $M^{}_{\rm R} = {\rm diag}\{M^{}_1, M^{}_2, M^{}_3\} \equiv \widehat{M}^{}_{\rm R}$ are diagonal, the neutrino mass spectrum and lepton flavor mixing are determined by the effective neutrino mass matrix $M^{}_\nu = - M^{}_{\rm D} \widehat{M}^{-1}_{\rm R} M^{\rm T}_{\rm D}$, which can be diagonalized as $M^{}_\nu = U \cdot {\rm diag}\{m^{}_1, m^{}_2, m^{}_3\} \cdot U^{\rm T}$ with $U$ being the Pontecorvo-Maki-Nakagawa-Sakata (PMNS) mixing matrix~\cite{Pontecorvo:1957cp, Maki:1962mu, Pontecorvo:1967fh}. Therefore, in order to obtain any predictions for the low-energy observables, one has to know the flavor structure of $M^{}_{\rm D}$, which is completely unconstrained in the generic seesaw model. Generally speaking, there are two different guiding principles towards seeking a solution to this problem, namely, flavor symmetry and minimality:
\begin{itemize}
\item In the first approach, discrete or continuous flavor symmetries are imposed on the generic Lagrangian in Eq.~(\ref{eq: massterm}), and all the SM fields are assigned into proper representations of the symmetry groups. Due to the required symmetries, the Yukawa coupling matrices are not arbitrary any more. It has been demonstrated that discrete flavor symmetries can be implemented to successfully predict interesting lepton flavor mixing patterns, which are well compatible with the latest neutrino oscillation data. For recent reviews on this topic, see Refs.~\cite{Altarelli:2010gt, Ishimori:2010au, King:2013eh}. Although this scenario is very attractive in the first place, it actually suffers from the involvement of many new scalar fields that are needed in order to achieve the desired flavor structures of Yukawa coupling matrices. As a consequence, it is generally difficult to verify or disprove a flavor-symmetry model experimentally.

\item In the second approach, the number of model parameters is intentionally reduced to a level, beyond which the model would immediately run into contradictions with current experimental observations. The minimality of a model, in the sense of minimal number of free parameters, can be regarded as an Occam's razor~\cite{Harigaya:2012bw, Ohlsson:2012qh}. One practical way of reducing free parameters is to simply take some Yukawa matrix elements to be zero. The physical essence of texture zeros actually reflects that some elements in a Yukawa coupling matrix are highly suppressed when compared to the other elements, or they are irrelevant to fermion mass spectra and flavor mixing. For instance, the texture zeros turn out to be very useful to establish a relationship between small flavor mixing angles and strong mass hierarchy in the quark sector~\cite{Weinberg:1977hb, Fritzsch:1977za, Fritzsch:1977vd}. As shown by Weinberg in Ref.~\cite{Weinberg:1977hb}, the texture zeros in two-generation quark mass matrices lead to a successful prediction for the Cabbibo angle $\theta^{}_{\rm C} = \sqrt{m^{}_{\rm d}/m^{}_{\rm s}} \approx 0.22$, where the running mass of down quark $m^{}_{\rm d} = 2.82~{\rm MeV}$ and strange quark $m^{}_{\rm s} = 57~{\rm MeV}$ are evaluated at $M^{}_Z = 91.2~{\rm GeV}$~\cite{Xing:2011aa,Xing:2007fb}. In the same spirit, more than ten years ago, Frampton, Glashow and Yanagida proposed a minimal scenario of seesaw models, in which only two right-handed neutrinos are introduced and two elements of the Dirac neutrino mass matrix $M^{}_{\rm D}$ are assumed to be vanishing~\cite{Frampton:2002qc}. In this case, $M^{}_{\rm D}$ becomes a $3\times 2$ matrix, and can be explicitly written as
    \begin{equation}
    M^{}_{\rm D} = \left(\begin{matrix} {\bf 0}& a \cr a^\prime & {\bf 0} \cr b^\prime & b \end{matrix}\right) \; ,
    \label{eq: mD}
    \end{equation}
    where $a$, $b$, $a^\prime$ and $b^\prime$ are in general complex. There are totally fifteen possible patterns of $M^{}_{\rm D}$ with two texture zeros in different positions, and we shall examine all of them in the following section. The number of texture zeros in $M^{}_{\rm D}$ cannot be further increased, otherwise the model will be in conflict with three nonzero flavor mixing angles, as measured in neutrino oscillation experiments~\cite{Frampton:2002qc, Guo:2003cc, Mei:2003gn, Guo:2006qa}. On the other hand, the seesaw model with just one heavy right-handed neutrino does not work, since there will be two massless ordinary neutrinos that have already been excluded. Hence, the scenario of two heavy right-handed neutrinos together with the Frampton-Glashow-Yanagida (FGY) ansatz like that in Eq.~(\ref{eq: mD}) is the minimal version of type-I seesaw model, which will be called the FGY model hereafter. One can immediately verify that neutrino mass spectrum and leptonic CP-violating phases are calculable from the observed three neutrino mixing angles and two neutrino mass-squared differences~\cite{Guo:2006qa}, implying a complete testability of the model in future neutrino experiments. It is worthwhile to stress that this minimal scenario emerges when one right-handed Majornana neutrino is much heavier than the other two and decouples from the theory, or its Yukawa couplings to lepton and Higgs doublets are vanishingly small~\cite{Harigaya:2012bw}.
\end{itemize}
Another salient feature of the canonical seesaw model is to account for the baryon number asymmetry in our Universe via the leptogenesis mechanism~\cite{Fukugita:1986hr}. In the early Universe, the temperature is high enough to thermally produce heavy Majorana neutrinos $N^{}_i$. As the Universe cools down, the out-of-equilibrium and CP-violating decays of $N^{}_i$ generate lepton number asymmetries, which will further be converted into the baryon asymmetry via nonperturbative sphaleron processes~\cite{Manton:1983nd, Klinkhamer:1984di}. Excellent reviews on leptogenesis can be found in Refs.~\cite{Buchmuller:2004nz, Buchmuller:2005eh, Davidson:2008bu}.

In light of recent progress in neutrino oscillation experiments, we reconsider the FGY model and carry out a complete study with a focus on the currently unresolved problems, such as neutrino mass ordering, leptonic CP violation and the Majorana nature of neutrinos. The main motivation for such an investigation is two-fold. First, due to a minimal set of free parameters, the FGY model is quite predictive, so it is interesting to confront it with the latest global-fit results of neutrino oscillation data. A similar analysis has actually been done in Ref.~\cite{Harigaya:2012bw}. Different from that work, we take into account the renormalization-group (RG) running effects of lepton flavor mixing parameters from the seesaw scale $\Lambda^{}_{\rm SS}$, usually characterized by the lightest heavy Majorana neutrino mass $M^{}_1$, to the electroweak scale $\Lambda^{}_{\rm EW}$. Second, in the previous work, a strong mass hierarchy $M^{}_2 \gg M^{}_1$ is always assumed, and a narrow range of heavy neutrino masses $M^{}_1 \sim 5\times 10^{13}~{\rm GeV}$ is derived by requiring a successful leptogenesis mechanism to explain the cosmological matter-antimatter asymmetry. But such a large mass scale in the theory causes the naturalness or fine-tuning problem on the one hand~\cite{Vissani:1997ys,Abada:2007ux,Xing:2009in,Farina:2013mla,Clarke:2015gwa}, and the gravitino overproduction problem if the model is supersymmetrized on the other hand~\cite{Giudice:2003jh}. Therefore, we are motivated to go beyond the hierarchical limit, and consider both mild mass hierarchy and a nearly-degenerate mass spectrum of heavy Majorana neutrinos. Only with careful studies of RG running effects and general mass spectra of heavy Majorana neutrinos can we really test the FGY model.

The remaining part of our paper is organized as follows. In Section 2, phenomenological implications of the FGY model are explored and confronted with current neutrino oscillation data. We also consider the RG running effects of neutrino mixing parameters, and specify the allowed regions of the parameter space at the low-energy scale. Only four out of fifteen patterns of the Dirac neutrino Yukawa coupling matrices are found to be compatible with neutrino oscillation data, and only the inverted neutrino mass ordering is allowed. Section 3 is devoted to the generation of baryon number asymmetry via leptogenesis, where we also discuss the impact of lepton flavor effects and non-hierarchical mass spectrum of heavy Majorana neutrinos. The flavor structure of four viable patterns leads to a non-vanishing CP asymmetry in one specific lepton flavor. We point out that a nearly-degenerate mass spectrum of heavy Majorana neutrinos is required to explain the baryon number asymmetry, and simultaneously avoid huge radiative corrections to the light Higgs boson mass. Finally, we summarize our main conclusions in Section 4.

\section{Neutrino Masses and Flavor Mixing}

We start with neutrino mass spectrum and flavor mixing parameters in the type-I seesaw model with only two right-handed heavy Majorana neutrinos. After some general remarks, we proceed to introduce the FGY ansatz and explore its phenomenological implications. The RG evolution of neutrino masses and mixing parameters is considered when we confront the FGY ansatz with low-energy neutrino oscillation data. Finally, the model parameters relevant for leptogenesis at the high-energy scale are determined.

\subsection{General Remarks}

In the basis where both the charged-lepton mass matrix $M^{}_l$ and the heavy Majorana neutrino mass matrix $M^{}_{\rm R}$ are diagonal, the diagonalization of the light neutrino mass matrix $M^{}_\nu = - M^{}_{\rm D} \widehat{M}^{-1}_{\rm R} M^{\rm T}_{\rm D}$ via $M^{}_\nu = U \widehat{M}^{}_\nu U^{\rm T}$ gives us neutrino mass eigenvalues $\widehat{M}^{}_\nu = {\rm diag}\{m^{}_1, m^{}_2, m^{}_3\}$ and the PMNS matrix $U$. Since only two right-handed neutrinos are introduced and their mass matrix $\widehat{M}^{}_{\rm R}$ is of rank two, it is straightforward to verify that the rank of effective neutrino mass matrix $M^{}_\nu$ is two. As a consequence, the lightest neutrino must be massless. In the case of normal mass ordering (NO) with $m^{}_1 = 0$, we get $m^{}_2 = \sqrt{\Delta m^2_{21}}$ and $m^{}_3 = \sqrt{\Delta m^2_{31}}$. In the case of inverted mass ordering (IO) with $m^{}_3 = 0$, we have $m^{}_1 = \sqrt{|\Delta m^2_{32}| - \Delta m^2_{21}}$ and $m^{}_2 = \sqrt{|\Delta m^2_{32}|}$. The neutrino mass-squared differences $\Delta m^2_{21} \equiv m^2_2 - m^2_1$ and $\Delta m^2_{31} \equiv m^2_3 - m^2_1$ (or $\Delta m^2_{32} \equiv m^2_3 - m^2_2$) can be measured in neutrino oscillation experiments in the case of NO (or IO). At present, however, it is unclear whether neutrino mass ordering is NO or IO. The ongoing long-baseline accelerator experiments T2K~\cite{Itow:2001ee} and NO$\nu$A~\cite{Ayres:2004js}, the forthcoming medium-baseline reactor experiments JUNO~\cite{Li:2013zyd} and RENO-50~\cite{Park:2014sja}, and the future huge atmospheric neutrino experiment PINGU~\cite{Aartsen:2014oha} will provide a definitive answer to this question.

Furthermore, the PMNS matrix in this minimal model can be parametrized via three mixing angles $\{\theta^{}_{12}, \theta^{}_{13}, \theta^{}_{23}\}$, one Dirac-type CP-violating phase $\delta$ and one Majorana-type CP-violating phase $\sigma$, namely
\begin{eqnarray}
U = \left( \begin{matrix} c^{}_{12} c^{}_{13} & s^{}_{12} c^{}_{13}
& s^{}_{13} e^{-{\rm i} \delta} \cr -s^{}_{12} c^{}_{23} - c^{}_{12} s^{}_{13} s^{}_{23} e^{{\rm i} \delta} & + c^{}_{12} c^{}_{23} - s^{}_{12} s^{}_{13} s^{}_{23} e^{{\rm i} \delta} & c^{}_{13} s^{}_{23} \cr + s^{}_{12} s^{}_{23} - c^{}_{12} s^{}_{13} c^{}_{23} e^{{\rm i} \delta} & - c^{}_{12} s^{}_{23} - s^{}_{12} s^{}_{13} c^{}_{23} e^{{\rm i} \delta} & c^{}_{13} c^{}_{23} \cr \end{matrix} \right) \left(\begin{matrix} 1 & 0 & 0 \cr 0 & e^{i\sigma} & 0 \cr 0 & 0 & 1\end{matrix}\right) \; ,
\label{eq: PMNS}
\end{eqnarray}
where $c^{}_{ij} \equiv \cos \theta^{}_{ij}$ and $s^{}_{ij} \equiv \sin \theta^{}_{ij}$ have been defined for $ij = 12, 13, 23$. While three mixing angles have been determined with reasonably good precision from oscillation experiments, there is still no significant evidence for a nontrivial Dirac CP-violating phase. In Table 1, the latest global-fit analysis of neutrino oscillation parameters has been presented. One can observe that the best-fit value of Dirac CP-violating phase is $\delta = 306^\circ$ for NO and $\delta = 254^\circ$ for IO, but it becomes arbitrary at the $3\sigma$ level. The proposed neutrino super-beam experiments and neutrino factories are able to probe $\delta$ down to a few degrees~\cite{NF}.
\begin{table}[t]
\begin{center}
\vspace{-0.25cm} \caption{The best-fit values, together with the
1$\sigma$, 2$\sigma$ and 3$\sigma$ intervals, for three neutrino
mixing angles $\{\theta^{}_{12}, \theta^{}_{13}, \theta^{}_{23}\}$, two mass-squared differences $\{\Delta m^2_{21}, \Delta m^2_{31}~{\rm or}~\Delta m^2_{32}\}$ and the Dirac CP-violating phase $\delta$ from a global analysis of current experimental data~\cite{Gonzalez-Garcia}. Two independent global-fit analyses can be found in Refs.~\cite{Fogli,Valle}, which are in perfect agreement with the results presented here at the $3\sigma$ level.} \vspace{0.2cm}
\begin{tabular}{c|c|c|c|c}
\hline
\hline
Parameter & Best fit & 1$\sigma$ range & 2$\sigma$ range & 3$\sigma$ range \\
\hline
\multicolumn{5}{c}{Normal neutrino mass ordering
$(m^{}_1 < m^{}_2 < m^{}_3$)} \\ \hline
$\theta_{12}/^\circ$
& $33.48$ & 32.73 --- 34.26 & 31.98 --- 35.04 & 31.29 --- 35.91 \\
$\theta_{13}/^\circ$
& $8.50$ & 8.29 --- 8.70 & 8.08 --- 8.90 & 7.85 --- 9.10 \\
$\theta_{23}/^\circ$
& $42.3$  & 40.7 --- 45.3 & 39.1 --- 48.3 & 38.2 --- 53.3 \\
$\delta/^\circ$ &  $306$ & 236 --- 345 & 0 --- 24
$\oplus$ 166 --- 360 & 0 --- 360 \\
$\Delta m^2_{21}/[10^{-5}~{\rm eV}^2]$ &  $7.50$ & 7.33 --- 7.69 & 7.16 --- 7.88 & 7.02 --- 8.09 \\
$\Delta m^2_{31}/[10^{-3}~{\rm eV}^2]$ &  $+2.457$ & +2.410 --- +2.504 & +2.363 --- +2.551 & +2.317 --- +2.607 \\\hline
\multicolumn{5}{c}{Inverted neutrino mass ordering
$(m^{}_3 < m^{}_1 < m^{}_2$)} \\ \hline
$\theta_{12}/^\circ$
& $33.48$ & 32.73 --- 34.26 & 31.98 --- 35.04 & 31.29 --- 35.91 \\
$\theta_{13}/^\circ$
& $8.51$ & 8.30 --- 8.71 & 8.09 --- 8.91 & 7.87 --- 9.11 \\
$\theta_{23}/^\circ$
& $49.5$  & 47.3 --- 51.0 & 45.1 --- 52.5 & 38.6 --- 53.3 \\
$\delta/^\circ$ &  $254$ & 192 --- 317 & 0 --- 20
$\oplus$ 130 --- 360 & 0 --- 360 \\
$\Delta m^2_{21}/[10^{-5}~{\rm eV}^2]$ &  $7.50$ & 7.33 --- 7.69 & 7.16 --- 7.88 & 7.02 --- 8.09 \\
$\Delta m^2_{32}/[10^{-3}~{\rm eV}^2]$ &  $-2.449$ & $-2.496$ --- $-2.401$ & $-2.543$ --- $-2.355$ & $-2.590$ --- $-2.307$ \\ \hline\hline
\end{tabular}
\label{tb: gfit}
\end{center}
\end{table}

Since there is one massless neutrino, we have only one Majorana CP-violating phase $\sigma$. The observation of neutrinoless double-beta decays is the unique and feasible way to establish that neutrinos are Majorana particles, i.e., they are their own antiparticles~\cite{Rodejohann:2011mu}. The decay rate depends on the effective neutrino mass defined as $m^{}_{\beta \beta} \equiv |U^2_{e1} m^{}_1 + U^2_{e2} m^{}_2 + U^2_{e3} m^{}_3|$, where $U^{}_{ei}$ for $i=1, 2, 3$ denote the elements in the first row of the PMNS matrix $U$. More explicitly,
\begin{eqnarray}
m^{}_{\beta \beta} = \left\{\begin{tabular}{lcl}
                              $\sqrt{\Delta m^2_{31}} \cos^2 \theta^{}_{13} \left[\xi^2 \sin^4 \theta^{}_{12} + \tan^4 \theta^{}_{13} + 2\xi \sin^2 \theta^{}_{12} \tan^2 \theta^{}_{13} \cos 2(\sigma + \delta)\right]^{1/2}$ & ~ & for NO \;; \\
                              ~ & ~ & ~ \\
                              $\sqrt{|\Delta m^2_{32}|} \cos^2 \theta^{}_{13} \cos^2 \theta^{}_{12} \left[\zeta^2 + \tan^4 \theta^{}_{12} + 2\zeta \tan^2 \theta^{}_{12} \cos 2\sigma \right]^{1/2}$ & ~ & for IO \;, \\
                            \end{tabular}\right.
\label{eq: mbb}
\end{eqnarray}
where $\xi \equiv m^{}_2/m^{}_3$ and $\zeta \equiv m^{}_1/m^{}_2$. Now that neutrino masses are completely fixed by two mass-squared differences, we can get $\xi = \sqrt{\Delta m^2_{21}}/\sqrt{\Delta m^2_{31}} \approx 0.175$ and $\zeta = \sqrt{1 - \Delta m^2_{21}/|\Delta m^2_{32}|} \approx 0.985$ by using the best-fit values of neutrino mass-squared differences in Table 1. Notice that the relation $\xi^2 \approx 1 - \zeta^2 \approx \sqrt{2} \sin^2\theta^{}_{13} \approx 0.03$ holds as an excellent approximation. The exact value of $m^{}_{\beta \beta}$ depends on the Majorana CP-violating phase $\sigma$ in the IO case, and a combination of two unknown CP-violating phases $\sigma$ and $\delta$ in the NO case. However, it is straightforward to find out the lower and upper limits~\cite{Guo:2006mx, Xing:2014yka, Xing:2015zha}. For NO, we get
\begin{eqnarray}
\sqrt{\Delta m^2_{31}} \cos^2 \theta^{}_{13} \left(\xi \sin^2 \theta^{}_{12} - \tan^2 \theta^{}_{13} \right) \leq m^{}_{\beta \beta} \leq \sqrt{\Delta m^2_{31}} \cos^2 \theta^{}_{13} \left(\xi \sin^2 \theta^{}_{12} + \tan^2 \theta^{}_{13} \right) \; ,
\label{eq: mbbNO}
\end{eqnarray}
leading to $m^{}_{\beta \beta} \in [1.5, 3.7]~{\rm meV}$ with the help of the best-fit values in Table 1. For IO, we arrive at
\begin{eqnarray}
\sqrt{|\Delta m^2_{32}|} \cos^2 \theta^{}_{13} \cos^2 \theta^{}_{12} \left(\zeta - \tan^2 \theta^{}_{12} \right) \leq m^{}_{\beta \beta} \leq \sqrt{|\Delta m^2_{32}|} \cos^2 \theta^{}_{13} \cos^2 \theta^{}_{12} \left(\zeta + \tan^2 \theta^{}_{12} \right) \; ,
\label{eq: mbbIO}
\end{eqnarray}
implying $m^{}_{\beta \beta} \in [18, 48]~{\rm meV}$ with the best-fit values as inputs. As the future neutrinoless double-beta decay experiments are able to reach a sensitivity of about $20~{\rm meV}$~\cite{Rodejohann:2011mu}, the IO case seems to be more encouraging and phenomenologically interesting. Moreover, in this minimal seesaw model, the observation of neutrinoless double-beta decays may also pin down the unique Majorana CP-violating phase $\sigma$ via Eq.~(\ref{eq: mbb}), as long as the other mixing parameters can be well measured in neutrino oscillation experiments.

\subsection{The Frampton-Glashow-Yanagida Ansatz}

Although neutrino mass spectrum can be fixed by the observed neutrino mass-squared differences in the minimal seesaw model, three mixing angles and two CP-violating phases are in general arbitrary. Further restrictions on the flavor structure can induce testable correlations among low-energy observables. In the full theory above the seesaw scale $\Lambda^{}_{\rm SS}$, relevant parameters are the Dirac neutrino Yukawa coupling matrix $Y^{}_\nu$ and heavy Majorana neutrino masses $\{M^{}_1, M^{}_2\}$. If two elements of $Y^{}_\nu$ are vanishing~\cite{Frampton:2002qc}, there are fifteen logically possible patterns, which can be categorized into three classes:
\begin{itemize}
\item {\bf Case A} -- Two texture zeros are located in the same row, namely, $\left(Y^{}_\nu\right)^{}_{\alpha i} = \left(Y^{}_\nu\right)^{}_{\alpha j} = 0$ with $i \neq j$. There are only three patterns:
   \begin{eqnarray}
    \begin{tabular}{ccc}
      ${\bf A}^{}_1: \left(\begin{matrix} {\bf 0}& {\bf 0} \cr \times & \times \cr \times & \times \end{matrix}\right)$ \; , & ${\bf A}^{}_2: \left(\begin{matrix} \times & \times \cr {\bf 0} & {\bf 0} \cr \times & \times \end{matrix}\right)$ \; , & ${\bf A}^{}_3: \left(\begin{matrix} \times & \times \cr \times & \times \cr {\bf 0} & {\bf 0} \end{matrix}\right) $ \; , \\
    \end{tabular}
    \label{eq: CaseA}
  \end{eqnarray}
  where the cross `$\times$' denotes a nonzero matrix element.

\item {\bf Case B} -- Two texture zeros are located in different columns and rows, namely, $\left(Y^{}_\nu\right)^{}_{\alpha i} = \left(Y^{}_\nu\right)^{}_{\beta j} = 0$ with $\alpha \neq \beta$ and $i \neq j$. There are six patterns:
   \begin{eqnarray}
    \begin{tabular}{ccc}
      ${\bf B}^{}_1: \left(\begin{matrix} {\bf 0}& \times \cr \times & {\bf 0} \cr \times & \times \end{matrix}\right)$ \; , & ${\bf B}^{}_2: \left(\begin{matrix} {\bf 0}& \times \cr \times & \times\cr \times & {\bf 0} \end{matrix}\right)$ \; , & ${\bf B}^{}_3: \left(\begin{matrix} \times & \times \cr {\bf 0} & \times \cr \times & {\bf 0} \end{matrix}\right) $ \; , \\
      ~ & ~ & ~ \\
      ${\bf B}^{}_4: \left(\begin{matrix} \times & {\bf 0} \cr {\bf 0} & \times \cr \times & \times \end{matrix}\right)$ \; , & ${\bf B}^{}_5: \left(\begin{matrix} \times & {\bf 0} \cr \times & \times\cr {\bf 0} & \times \end{matrix}\right)$ \; , & ${\bf B}^{}_6: \left(\begin{matrix} \times & \times \cr \times & {\bf 0} \cr {\bf 0} & \times \end{matrix}\right) $ \; , \\
    \end{tabular}
    \label{eq: CaseB}
  \end{eqnarray}
  where the patterns ${\bf B}^{}_{4, 5, 6}$ are derived from ${\bf B}^{}_{1, 2, 3}$ by exchanging two columns.

\item {\bf Case C} -- Two texture zeros are located in the same column, namely, $\left(Y^{}_\nu\right)^{}_{\alpha i} = \left(Y^{}_\nu\right)^{}_{\beta i} = 0$ with $\alpha \neq \beta$. There are six patterns:
   \begin{eqnarray}
    \begin{tabular}{ccc}
      ${\bf C}^{}_1: \left(\begin{matrix} {\bf 0}& \times \cr {\bf 0} & \times\cr \times & \times \end{matrix}\right)$ \; , & ${\bf C}^{}_2: \left(\begin{matrix} {\bf 0}& \times \cr \times & \times\cr {\bf 0} & \times\end{matrix}\right)$ \; , & ${\bf C}^{}_3: \left(\begin{matrix} \times & \times \cr {\bf 0} & \times \cr {\bf 0} & \times \end{matrix}\right) $ \; , \\
      ~ & ~ & ~ \\
      ${\bf C}^{}_4: \left(\begin{matrix}\times & {\bf 0}\cr \times &{\bf 0} \cr \times & \times \end{matrix}\right)$ \; , & ${\bf C}^{}_5: \left(\begin{matrix} \times & {\bf 0} \cr \times & \times \cr \times & {\bf 0} \end{matrix}\right)$ \; , & ${\bf C}^{}_6: \left(\begin{matrix} \times & \times \cr \times & {\bf 0} \cr \times & {\bf 0} \end{matrix}\right) $ \; , \\
    \end{tabular}
    \label{eq: CaseC}
  \end{eqnarray}
 where the patterns ${\bf C}^{}_{4, 5, 6}$ can be obtained from ${\bf C}^{}_{1, 2, 3}$ by exchanging two columns.
\end{itemize}
It is worth pointing out that the patterns in each class can be related by the elementary transformations, i.e., the $3\times 3$ elementary matrices ${\cal P}^{}_{ij}$ (for $ij = 12, 23, 13$) and the $2\times 2$ elementary matrix ${\cal Q}$. The action of ${\cal P}^{}_{ij}$ from left (or right) induces an exchange between $i$-th and $j$-th rows (or columns), and likewise for ${\cal Q}$. With the help of ${\cal P}^{}_{ij}$ and ${\cal Q}$, one can change the positions of texture zeros. For instance, we have $Y^{}_\nu({\bf A}^{}_2) = {\cal P}^{}_{12} Y^{}_\nu({\bf A}^{}_1)$ and $Y^{}_\nu({\bf A}^{}_3) = {\cal P}^{}_{13} Y^{}_\nu({\bf A}^{}_1)$. In a similar way, one can prove that all the patterns in {\bf Case B} can be obtained from $Y^{}_\nu({\bf B}^{}_1)$ by using the elementary transformations. To be explicit, we list the relevant relations
\begin{eqnarray}
\begin{tabular}{ccccc}
$Y^{}_\nu ({\bf B}^{}_2) = {\cal P}^{}_{23} Y^{}_\nu ({\bf B}^{}_1)$ \; , & ~ & $Y^{}_\nu ({\bf B}^{}_3) = {\cal P}^{}_{12} {\cal P}^{}_{23} Y^{}_\nu ({\bf B}^{}_1)$ \; , & ~ & $Y^{}_\nu({\bf B}^{}_{i+3}) = Y^{}_\nu({\bf B}^{}_{i}) {\cal Q}$ \; ,
\end{tabular}
\label{eq: transf}
\end{eqnarray}
where the index $i = 1, 2, 3$ in the last equality is implied. The same transformations apply to the patterns in Eq.~(\ref{eq: CaseC}). As we will show later in this section, the above observations will be useful to analyze the texture zeros in the effective neutrino mass matrix $M^{}_\nu$. Note that the elementary transformations are implemented to examine the location of texture zeros, so the nonzero elements in both $Y^{}_\nu$ and the corresponding $M^{}_\nu$ are not necessarily identical for each pattern.

Below the seesaw scale, one can integrate out heavy Majorana neutrinos and obtain the unique Weinberg operator ${\cal O}^{}_5 =  (\kappa/2) ~ (\overline{\ell^{}_{\rm L}} \tilde{H}) \cdot (\tilde{H}^{\rm T}\ell^{\rm c}_{\rm L})$ of dimension five~\cite{Weinberg:1979sa} with $\kappa = - Y^{}_\nu \widehat{M}^{-1}_{\rm R} Y^{\rm T}_\nu$. After the spontaneous gauge symmetry breaking, neutrinos acquire tiny Majorana masses from the Weinberg operator and their mass matrix is $M^{}_\nu = \kappa v^2$, which is just the seesaw formula in the language of effective theories. Now it is clear that $Y^{}_\nu$ is given at a superhigh-energy scale $\mu = M^{}_1$, but neutrino oscillation parameters are measured at low energies. In order to study whether the flavor structure of $Y^{}_\nu$ in {\bf Case A}, {\bf B} and {\bf C} is viable, we have to examine the RG evolution of $\kappa$ from the seesaw scale $\Lambda^{}_{\rm SS}$ to the electroweak scale $\Lambda^{}_{\rm EW}$, and compare the predictions from $\kappa(\Lambda^{}_{\rm EW})$ with neutrino oscillation data.

Given $Y^{}_\nu$ in Eqs.~(\ref{eq: CaseA})--(\ref{eq: CaseC}), we are ready to check if $\kappa$ inherits some texture zeros from $Y^{}_\nu$. Since all the patterns in each class are related by ${\cal P}^{}_{ij}$ and ${\cal Q}$ matrices, it is sufficient to consider the first pattern and perform the corresponding elementary transformations to derive the results for the others. More explicitly, we have $\kappa(M^{}_1)$ at the seesaw scale
\begin{eqnarray}
\begin{tabular}{ccccc}
$\kappa^{}_{{\bf A}^{}_1}: \left(\begin{matrix} {\bf 0} & {\bf 0} & {\bf 0} \cr {\bf 0} & \times & \times \cr {\bf 0} & \times & \times \end{matrix}\right)$ \; , & ~ & $\kappa^{}_{{\bf B}^{}_1}: \left(\begin{matrix} \times & {\bf 0} & \times \cr {\bf 0} & \times & \times \cr \times & \times & \times \end{matrix}\right)$ \; , & ~ & $\kappa^{}_{{\bf C}^{}_1}: \left(\begin{matrix} \times & \times & \times \cr \times & \times & \times \cr \times & \times & \times \end{matrix}\right)$ \; , \\
\end{tabular}
\label{eq: kappa}
\end{eqnarray}
where one can observe that the patterns ${\bf C}^{}_i$ (for $i = 1, 2, \cdots, 6$) do not lead to any texture zeros in $\kappa$. For {\bf Case A} in Eq.~(\ref{eq: CaseA}), it is easy to derive $\kappa^{}_{{\bf A}^{}_j} = {\cal P}^{}_{1j} \kappa^{}_{{\bf A}^{}_1} {\cal P}^{}_{1j}$ for $i = 2, 3$, so $\kappa$ in this case has a nonzero $2\times 2$ block submatrix. For {\bf Case B} in Eq.~(\ref{eq: CaseB}), with the help of Eq.~(\ref{eq: transf}), we arrive at the following identities
\begin{eqnarray}
\begin{tabular}{ccccc}
$\kappa^{}_{{\bf B}^{}_2} = {\cal P}^{}_{23} \kappa^{}_{{\bf B}^{}_1} {\cal P}^{}_{23}$ \;, & ~ & $\kappa^{}_{{\bf B}^{}_3} = {\cal P}^{}_{12} {\cal P}^{}_{23} \kappa^{}_{{\bf B}^{}_1} {\cal P}^{}_{23} {\cal P}^{}_{12}$ \;, & ~ & $\kappa^{}_{{\bf B}^{}_{i+3}} = \kappa^{}_{{\bf B}^{}_i}$ \;, \\
\end{tabular}
\label{eq: transfkappa}
\end{eqnarray}
where the last identity indicates that one texture zero is located in the same position in $\kappa$ for ${\bf B}^{}_{i+3}$ and ${\bf B}^{}_{i}$ for $i = 1, 2, 3$.

\subsection{Renormalization-Group Running Effects}

As we have mentioned, neutrino masses at the sub-eV level indicate that the seesaw scale is extremely high $\Lambda^{}_{\rm SS} \sim 10^{14}~{\rm GeV}$, if the Dirac neutrino Yukawa couplings are of order ${\cal O}(1)$. In the full theory above the seesaw scale, two heavy Majorana neutrinos are added into the SM particle content, and they interact with the SM particles only through the Yukawa interaction, which is governed by the coupling matrix $Y^{}_\nu$. After taking into account radiative corrections and renormalizing the model in the scheme of dimensional regularization and modified minimal subtraction, we are left with coupling and mass parameters that depend on the renormalization scale $\mu$. The evolution of model parameters with respect to $\mu$ is described by their RG equations. For $\mu < \Lambda^{}_{\rm SS}$, the decoupling of heavy Majorana neutrinos is treated by explicitly integrating them out, and the low-energy effective theory turns out to be just the SM plus a dimension-five operator, which is responsible for neutrino masses. At the one-loop level, the RG running effects of neutrino masses and flavor mixing parameters can be studied by solving the RG equation of $\kappa$~\cite{Chankowski:1993tx, Babu:1993qv, Antusch:2001ck}
\begin{equation}
16\pi^2 \frac{{\rm d}\kappa}{{\rm d}t} = \alpha^{}_\kappa \kappa + C^{}_\kappa \left[ \left(Y^{}_l Y^\dagger_l\right) \kappa + \kappa \left(Y^{}_l Y^\dagger_l\right)^{\rm T}\right] \; ,
\label{eq: RGEkappa}
\end{equation}
with $t \equiv \ln (\mu/\Lambda^{}_{\rm EW})$. In the SM, the relevant coefficients in Eq.~(\ref{eq: RGEkappa}) are $C^{}_\kappa = -3/2$ and $\alpha^{}_\kappa \approx - 3 g^2_2 + 6 y^2_t + \lambda$, where $g^{}_2$ stands for the ${\rm SU(2)^{}_L}$ gauge coupling, $y^{}_t$ the top-quark Yukawa coupling, and $\lambda$ the Higgs self-coupling constant. If the dimension-five Weinberg operator is derived in the minimal supersymmetric standard model (MSSM), we have $M^{}_\nu = \kappa (v\sin \beta)^2$ with $\tan \beta$ being the ratio of vacuum expectation values of two MSSM Higgs doublets. In this framework, the RG equation of $\kappa$ is still given by Eq.~(\ref{eq: RGEkappa}) but with $C^{}_\kappa = 1$ and $\alpha^{}_\kappa \approx -6g^2_1/5 - 6g^2_2 + 6 y^2_t$. Note that only the top-quark Yukawa coupling is retained in $\alpha^{}_\kappa$, as the Yukawa couplings of other fermions are much smaller and have safely been neglected. The RG evolution of neutrino masses and lepton flavor mixing parameters has been extensively studied in the literature~\cite{Antusch:2003kp, Antusch:2005gp, Mei:2005qp, Xing:2005fw, Xing:2007fb, Luo:2012ce}. See, e.g., Ref.~\cite{Ohlsson:2013xva}, for a recent review on this topic.

Working in the basis where the charged-lepton Yukawa coupling matrix $Y^{}_l = {\rm diag}\{y^{}_e, y^{}_\mu, y^{}_\tau\}$ is diagonal, we can solve Eq.~(\ref{eq: RGEkappa}) and obtain
\begin{eqnarray}
\kappa(\Lambda^{}_{\rm EW}) = I^{}_0 \left(\begin{matrix} I^{}_e & 0 & 0 \cr 0 & I^{}_\mu & 0 \cr 0 & 0 & I^{}_\tau \end{matrix}\right) \kappa(M^{}_1) \left(\begin{matrix} I^{}_e & 0 & 0 \cr 0 & I^{}_\mu & 0 \cr 0 & 0 & I^{}_\tau \end{matrix}\right) \; ,
\label{eq: solkappa}
\end{eqnarray}
where the evolution functions read
\begin{eqnarray}
I^{}_0 &=& \exp \left[ - \frac{1}{16\pi^2} \int^{\ln(M^{}_1/\Lambda^{}_{\rm EW})}_0 \alpha^{}_\kappa(t) ~ {\rm d}t\right] \; , \\
\label{eq: Izero}
I^{}_\alpha &=& \exp \left[ - \frac{C^{}_\kappa}{16\pi^2} \int^{\ln(M^{}_1/\Lambda^{}_{\rm EW})}_0 y^2_\alpha(t) ~ {\rm d}t\right] \;,
\label{eq: Ialfa}
\end{eqnarray}
for $\alpha = e, \mu, \tau$. From Eq.~(\ref{eq: solkappa}), it is now evident how the low-energy observables residing in $M^{}_\nu = \kappa(\Lambda^{}_{\rm EW}) v^2$ are related to the model parameters in $\kappa(M^{}_1)$ at a high-energy scale. In the following, we show that
it is already possible to exclude most patterns in Eqs. (\ref{eq: CaseA})--(\ref{eq: CaseC}) based on the solution in Eq.~(\ref{eq: solkappa}).
\begin{enumerate}
\item An important observation from Eq.~(\ref{eq: solkappa}) is that texture zeros in $\kappa$ are rather stable against the RG running. On the other hand, Eq.~(\ref{eq: kappa}) tells us that $\kappa(M^{}_1)$ for the patterns ${\bf A}^{}_i$ possesses five vanishing elements, appearing in the $i$-th row and $i$-th column. Therefore, $\kappa(\Lambda^{}_{\rm EW})$ in {\bf Case A} inherits the same structure of $\kappa(M^{}_1)$, leading to just one nontrivial mixing angle, which has already been excluded by current neutrino oscillation data. Thus, all three patterns in Eq.~(\ref{eq: CaseA}) are ruled out.

\item Then we turn to the patterns ${\bf B}^{}_{1, 2, 3}$, and the same conclusions should also be applicable to ${\bf B}^{}_{4, 5, 6}$, since the texture zero in $M^{}_\nu$ is located in the same position. For this class, there is only one texture zero in $\kappa(\Lambda^{}_{\rm EW})$ or $M^{}_\nu = \kappa(\Lambda^{}_{\rm EW}) v^2$ in the off-diagonal position, namely,
    \begin{eqnarray}
    (M^{}_\nu)^{}_{\alpha \beta} = \sum_i m^{}_i U^{}_{\alpha i} U^{}_{\beta i} = 0 \; ,
    \label{eq: zero}
    \end{eqnarray}
    for $(\alpha, \beta) = (e, \mu)$, $(e, \tau)$ and $(\mu, \tau)$. When the RG running effects are considered, Eq.~(\ref{eq: solkappa}) indicates that the texture zero remains in the effective neutrino mass matrix $M^{}_\nu$. The constraints on neutrino masses and mixing matrix elements in Eq.~(\ref{eq: zero}) can be expressed as
    \begin{equation}
    \begin{tabular}{ccl}
    $U^{}_{\alpha 2} U^{}_{\beta 2} m^{}_2 + U^{}_{\alpha 3} U^{}_{\beta 3} m^{}_3 = 0$ \;  & ~ & for NO \; ,  \\
    $U^{}_{\alpha 1} U^{}_{\beta 1} m^{}_1 + U^{}_{\alpha 2} U^{}_{\beta 2} m^{}_2 = 0$ \;  & ~ & for IO \; , \\
    \end{tabular}
    \label{eq: const}
    \end{equation}
    which have been investigated in Ref.~\cite{Harigaya:2012bw}, where the latest neutrino oscillation data are implemented but the RG running effects are entirely ignored. In the NO case, it has been found that all the patterns in Eq.~(\ref{eq: CaseB}) are ruled out mainly due to the observed $\theta^{}_{13}$~\cite{An:2012eh, An:2013uza, An:2013zwz, Ahn:2012nd}. In the IO case, $(M^{}_\nu)^{}_{\mu \tau} = 0$ is shown to be strongly disfavored, so the patterns ${\bf B}^{}_3$ and ${\bf B}^{}_6$ are excluded. Hence, according to Ref.~\cite{Harigaya:2012bw}, only ${\bf B}^{}_{1, 2}$ and ${\bf B}^{}_{4, 5}$ in the IO case are compatible with the latest neutrino oscillation data.

\item Since the patterns in Eq.~(\ref{eq: CaseC}) do not imply any zero elements in $\kappa(M^{}_1)$, the analysis of {\bf Case C} in Ref.~\cite{Harigaya:2012bw} seems to be not applicable. Thus it is expected the predictions at a superhigh-energy scale will be significantly changed at the low-energy scale. However, as we demonstrate below, a characteristic relationship among the elements in $\kappa$ is maintained at the low-energy scale and validates the conclusions in Ref.~\cite{Harigaya:2012bw}. Let us take the pattern ${\bf C}^{}_1$ for example, and specify its matrix elements:
    \begin{eqnarray}
    \begin{tabular}{ccc}
    ${\bf C}^{}_1 : \left(\begin{matrix} {\bf 0} & a \cr {\bf 0} & b \cr a^\prime & b^\prime \end{matrix}\right)$ \; , & ~ & $\kappa(M^{}_1) = \displaystyle \frac{1}{M^{}_1} \left(\begin{matrix} 0 & 0 & 0 \cr 0 & 0 & 0 \cr 0 & 0 & {a^\prime}^2 \end{matrix}\right) + \frac{1}{M^{}_2} \left(\begin{matrix} a^2 & ab & ab^\prime \cr ab & b^2 & bb^\prime \cr ab^\prime & bb^\prime & \displaystyle {b^\prime}^2 \end{matrix}\right) $ \; ,
    \end{tabular}
    \label{eq: newC}
    \end{eqnarray}
    where the corresponding $\kappa(M^{}_1)$ has been given as well. Combining Eq.~(\ref{eq: solkappa}) and Eq.~(\ref{eq: newC}), one can verify that the relation
    \begin{equation}
    \displaystyle \frac{(M^{}_\nu)^{}_{ee}}{(M^{}_\nu)^{}_{\mu e}} = \displaystyle \frac{(M^{}_\nu)^{}_{e\mu}}{(M^{}_\nu)^{}_{\mu \mu}} = \displaystyle \frac{(M^{}_\nu)^{}_{e\tau}}{(M^{}_\nu)^{}_{\mu \tau}}
    \label{eq: newrel}
    \end{equation}
    holds both for $\mu = \Lambda^{}_{\rm EW}$ and for $\mu = M^{}_1$. Therefore, it is adequate to inspect if the relationship in Eq.~(\ref{eq: newrel}) is satisfied by current neutrino oscillation data. More explicitly, the first identity in Eq.~(\ref{eq: newrel}) gives rise to $U^{}_{e 3} U^{}_{\mu 2} = U^{}_{e 2} U^{}_{\mu 3}$ for NO, and $U^{}_{e 2} U^{}_{\mu 1} = U^{}_{e 1} U^{}_{\mu 2}$ for IO, while the second identity is fulfilled automatically. The constraints for the other patterns can be found in a similar way. Those relations among the PMNS matrix elements have also been derived in Ref.~\cite{Harigaya:2012bw}, although in a different manner, and used to exclude all the patterns in Eq.~(\ref{eq: CaseC}) in both NO and IO cases.
\end{enumerate}

In summary, we have proved that texture zeros or proportionality relations in $\kappa(M^{}_1)$ are not spoiled by the RG running effects, so they also exist in $\kappa(\Lambda^{}_{\rm EW})$ at the low-energy scale. Consequently, neutrino oscillation data can be directly implemented to rule out most patterns of $Y^{}_\nu$ with two texture zeros. It turns out that only ${\bf B}^{}_{1, 2}$ and ${\bf B}^{}_{4, 5}$ in Eq.~(\ref{eq: CaseB}) in the case of IO are consistent with experimental data, which generalizes the conclusions reached in Ref.~\cite{Harigaya:2012bw} to the situation including radiative corrections.

\subsection{Viable Patterns}

Now we are left with just four viable patterns, namely ${\bf B}^{}_{1, 2}$ and ${\bf B}^{}_{4, 5}$ in Eq.~(\ref{eq: CaseB}), and only the IO case is allowed. The latter indicates a sizable value of $m^{}_{\beta \beta}$, around $50~{\rm meV}$, and thus is quite encouraging for future experiments to search for neutrinoless double-beta decays. Although the RG running effects are unable to revive any patterns in the NO case, they do have significant impact on the allowed regions of model parameters, particularly in the MSSM with a large $\tan \beta$. Hence, in this subsection, we examine four viable patterns in more detail, and explore the favored parameter space.

As we have shown in the previous subsections, the effective neutrino mass matrix $M^{}_\nu$ at the low-energy scale in this case contains one texture zero, which sets two constraining relations on neutrino masses and mixing angles.  Since neutrino mass spectrum is completely fixed by the observed neutrino mass-squared differences, one can determine two CP-violating phases in terms of neutrino masses and three mixing angles. According to Eq.~(\ref{eq: transfkappa}), the two patterns in each pair of $\{{\bf B}^{}_1, {\bf B}^{}_4\}$ and $\{{\bf B}^{}_2, {\bf B}^{}_5\}$ are related by an exchange between two columns, so the location of texture zero in $M^{}_\nu$ is identical, indicating the same low-energy predictions. However, the model parameters in the full theory at the seesaw scale are different, as we shall show later. Using the second identity in Eq.~(\ref{eq: const}) for the case of $(\alpha, \beta) = (e, \mu)$, we obtain
\begin{eqnarray}
m^{}_1 c^{}_{12} (c^{}_{23} s^{}_{12} + c^{}_{12} s^{}_{23} s^{}_{13}e^{{\rm i}\delta}) - m^{}_2 s^{}_{12} (c^{}_{12} c^{}_{23} - s^{}_{12} s^{}_{23} s^{}_{13} e^{{\rm i}\delta})e^{2{\rm i}\sigma} = 0 \; ,
\label{eq: constB}
\end{eqnarray}
whose real and imaginary parts allow us to determine $\delta$ and $\sigma$ via
\begin{eqnarray}
\cos \delta &=& \frac{s^2_{12} c^2_{12} c^2_{23} (1 - \zeta^2) + s^2_{23} s^2_{13}(s^4_{12} - \zeta^2 c^4_{12})} {2 s^{}_{12} c^{}_{12} s^{}_{23} c^{}_{23} s^{}_{13} (s^2_{12} + \zeta^2 c_{12}^2)} \; ,
\label{eq: delta}
\\
\cos 2\sigma &=& \frac{ s^2_{12} c^2_{12} c^2_{23} (1 + \zeta^2) - s^2_{23} s^2_{13} ( s_{12}^4 + \zeta^2 c_{12}^4)}{2 \zeta s_{12}^2 c_{12}^2 (c_{23}^2 + s_{23}^2 s_{13}^2)} \; ,
\label{eq: sigma}
\end{eqnarray}
up to a sign ambiguity. Since $1 - \zeta^2 \approx \sqrt{2} s^2_{13} \approx 0.03$ holds as an excellent approximation, one can expand the right-hand sides of Eqs.~(\ref{eq: delta}) and (\ref{eq: sigma}) in terms of $1 - \zeta^2$ and $s^2_{13}$, and ignore the higher-order terms of ${\cal O}(s^3_{13})$. After a straightforward calculation, we arrive at
\begin{eqnarray}
\cos \delta &\approx& \frac{\sin 2\theta^{}_{12}}{4 \tan \theta^{}_{23} \sin \theta^{}_{13}} (1 - \zeta^2) - \frac{\tan \theta^{}_{23}}{\tan 2\theta^{}_{12}} \sin \theta^{}_{13} \; , \nonumber \\
\cos 2\sigma &\approx& 1 - \frac{\tan^2 \theta^{}_{23} \sin^2 \theta^{}_{13}}{2 \sin^2 \theta^{}_{12} \cos^2 \theta^{}_{12}} \; ,
\label{eq: approx}
\end{eqnarray}
implying that $\delta \approx 90^\circ$ or $270^\circ$ and $\sigma \approx 0^\circ$. The deviation of $\delta$ from the maximum $90^\circ$ or $270^\circ$, and that of $\sigma$ from zero, are on the order of $\theta^{}_{13}$ in the leading-order approximation. For the pattern ${\bf B}^{}_2$, one needs to consider Eq.~(\ref{eq: const}) with $(\alpha, \beta) = (e, \tau)$. It is easy to verify that Eqs.~(\ref{eq: delta})--(\ref{eq: approx}) become applicable to this case after replacing $\delta$ with $\delta + \pi$, as well as $\theta^{}_{23}$ with $\pi/2 - \theta^{}_{23}$, namely, flipping the octant of $\theta^{}_{23}$. This observation indicates that the determination of the octant of $\theta^{}_{23}$ and the measurement of CP-violating phases $\delta$ in future neutrino oscillation experiments can be used to distinguish between the patterns ${\bf B}^{}_1$ and ${\bf B}^{}_2$ for the Dirac neutrino Yukawa coupling matrix.

There are five real parameters in $M^{}_{\rm D}$, since two matrix elements are zero and three arbitrary phases can be absorbed by redefining the charged-lepton fields. Moreover, the heavy Majorana neutrino masses $M^{}_1$ and $M^{}_2$ are free parameters. It is convenient to introduce the Casas-Ibarra parametrization~\cite{Casas:2001sr}
\begin{equation}
M^{}_{\rm D} =  U \sqrt{\widehat{M}^{}_\nu} O \sqrt{\widehat{M}^{}_{\rm R}} = U \left(\begin{matrix} \sqrt{m^{}_1} & 0 & 0 \cr 0 & \sqrt{m^{}_2} & 0 \cr 0 & 0 & ~~0~ \end{matrix}\right) \left( \begin{matrix} \cos z & -\sin z \cr \sin z & \cos z \cr 0 & 0 \end{matrix}\right) \left( \begin{matrix} \sqrt{M^{}_1} & 0 \cr 0 & \sqrt{M^{}_2} \end{matrix} \right) \; ,
\label{eq: CI}
\end{equation}
where $U$ is the PMNS matrix given in Eq.~(\ref{eq: PMNS}), and $O$ is a $3\times 2$ orthogonal matrix with $z$ being a complex parameter, satisfying $O^{\rm T} O = O O^{\rm T} = {\bf 1}$. Note that we have concentrated on the IO case, which is the only allowed possibility in the FGY model. All the mixing angles, CP-violating phases, and neutrino masses in Eq.~(\ref{eq: CI}) should take values at the seesaw scale, which are in general distinct from those extracted from neutrino oscillation experiments at the low-energy scale (e.g., at the Fermi scale $M^{}_Z = 91.2~{\rm GeV}$). Because of the texture zeros in $M^{}_{\rm D}$, the CP-violating phases $\delta$ and $\sigma$ can be determined in terms of neutrino masses and mixing angles as in Eqs.~(\ref{eq: delta}) and (\ref{eq: sigma}), but now with their values at the seesaw scale. In addition, the complex parameter $z$ can be determined by
\begin{eqnarray}
\tan z = - \frac{U^{}_{e 1}}{U^{}_{e 2}} \sqrt{\frac{m^{}_1}{m^{}_2}} = - \frac{\sqrt{\zeta}}{\tan \theta^{}_{12}} e^{-{\rm i}\sigma} \; ,
\label{eq: tanz}
\end{eqnarray}
for ${\bf B}^{}_1$ and ${\bf B}^{}_2$. Since ${\bf B}^{}_4$ and ${\bf B}^{}_5$ are related to ${\bf B}^{}_1$ and ${\bf B}^{}_2$ by exchanging two columns, respectively, the parameter $z$ in the former two cases can be calculated first from Eq.~(\ref{eq: tanz}), and then followed by a shift of $z \to z + \pi/2$. Now it is evident that the complex parameter $z$ is actually determined by the neutrino mass ratio $\zeta = m^{}_1/m^{}_2$, the mixing angle $\theta^{}_{12}$ and the Majorana CP-violating phase $\sigma$. However, the RG running effects on these parameters, in particular $\theta^{}_{12}$ and $\sigma$, could be significant.
\begin{figure}[!t]
\begin{center}
\subfigure{%
\includegraphics[width=0.4\textwidth]{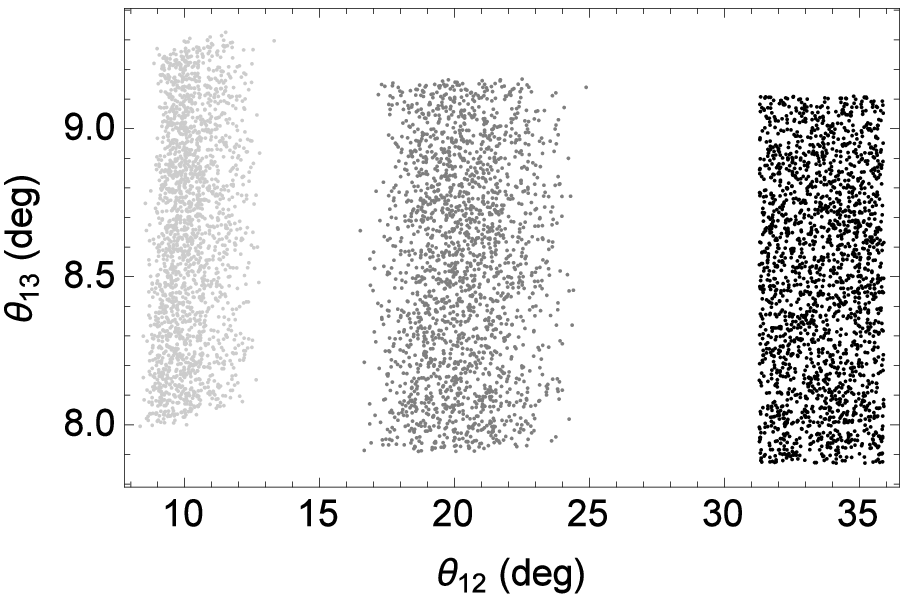}        }%
\subfigure{%
\hspace{0.5cm}
\includegraphics[width=0.4\textwidth]{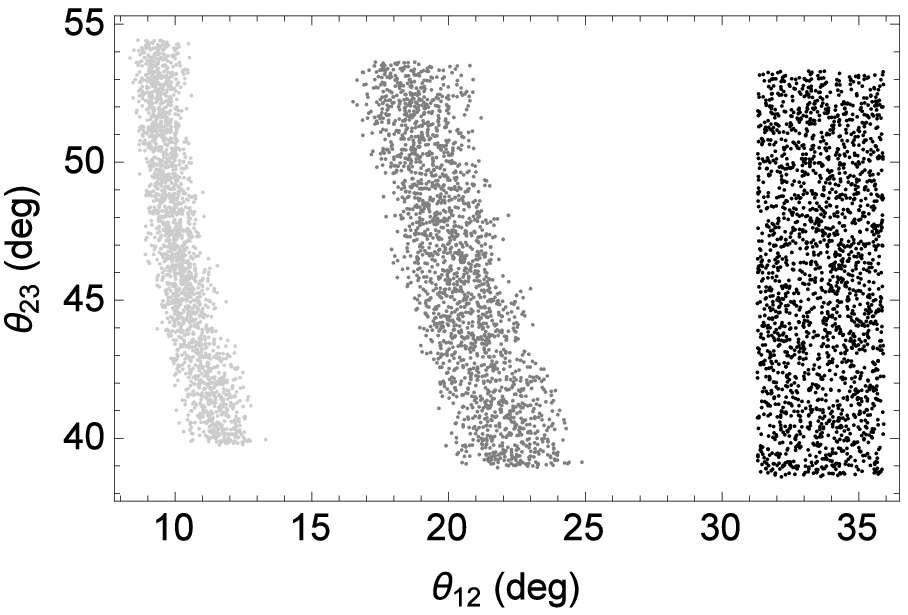}        } \\
\subfigure{%
\hspace{-0.0cm}
\includegraphics[width=0.4\textwidth]{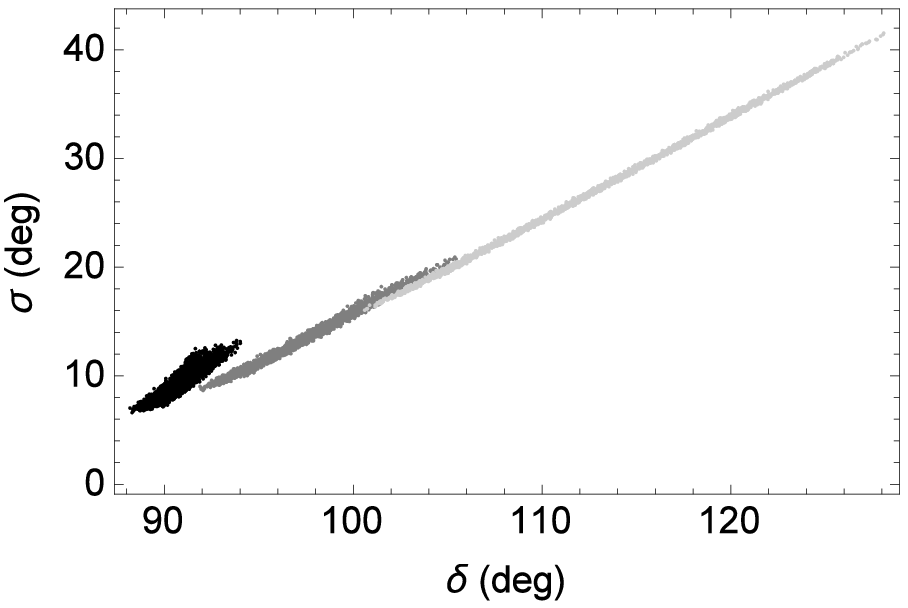}        }%
\subfigure{%
\hspace{0.45cm}
\includegraphics[width=0.4\textwidth]{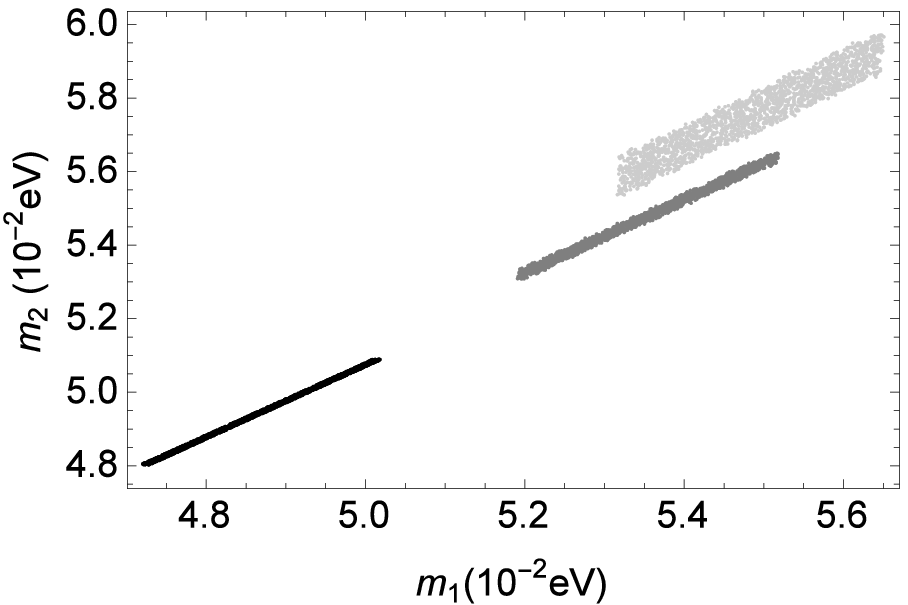}        }
\end{center}
\vspace{-0.3cm}
\caption{Illustration for the RG running effects on neutrino mixing angles $\{\theta^{}_{12}, \theta^{}_{13}, \theta^{}_{23}\}$, leptonic CP-violating phases $\{\delta, \sigma\}$ and neutrino masses $\{m^{}_1, m^{}_2\}$ for ${\bf Pattern ~ B}^{}_1$ in the MSSM, where the black points denote the parameters at $M^{}_Z = 91.2~{\rm GeV}$,  while the dark- and light-gray points represent the parameters at the seesaw scale $\Lambda^{}_{\rm SS} = 10^{13}~\rm{GeV}$ for $\tan\beta=30$ and $\tan\beta=50$, respectively. Note that $\delta$ and $\sigma$ also have another branch of solutions with their signs inverted simultaneously, and the mass scale of sparticles is taken to be $M^{}_{\rm SUSY} = 1~{\rm TeV}$.}
\label{fg:RGE_t3050}
\end{figure}

Taking ${\bf Pattern ~ B}^{}_1$ for example, we proceed to explore the possible parameter space at the low-energy scale by using the global-fit results in Table 1, and that at the high-energy scale by numerically solving the complete set of one-loop RG equations. In view of minimality of the FGY ansatz, we shall consider the minimal SM. In the SM, the largest charged-lepton Yukawa coupling $y_\tau^{}$ is as small as $10^{-2}$. According to Eq.~(\ref{eq: Ialfa}), the evolution function running from the electroweak scale to the seesaw scale $\Lambda_{\mathrm{SS}}^{} = 10^{13}~\mathrm{GeV}$ is approximately given by $I_\tau^{} \approx \mathrm{exp}(-25\times 10^{-6}) \approx 1$. Therefore, we have $I^{}_e \approx I^{}_\mu \approx I^{}_\tau \approx 1$, and the form of $\kappa$ remains unchanged during the RG running, resulting in negligible modifications on the mixing angles, CP-violating phases, and the ratio of neutrino masses. This means that the predictions of FGY ansatz are essentially valid at high-energy scales in the minimal SM. 

In the MSSM, the running effects are expected to be significant, since $y_\tau^{}$ can be enhanced by large values of $\tan\beta$. We first input the neutrino mixing angles and two neutrino mass-squared differences within their $3\sigma$ ranges at $M_Z^{}$. Two stages of RG running are then performed, namely, one from $M_Z^{}$ to the sparticle mass scale $M_{\mathrm{SUSY}}$ with the SM RG equations, and the other one from $M_{\mathrm{SUSY}}$ to $\Lambda_{\mathrm{SS}}^{} = 10^{13}~\mathrm{GeV}$ by adopting the MSSM RG equations. Taking $M_{\mathrm{SUSY}} = 1~\mathrm{TeV}$, we have calculated the running effects on neutrino mixing parameters, and the numerical results are presented in Fig.~\ref{fg:RGE_t3050}. We have also tried to vary this intermediate sparticle mass scale $M_{\mathrm{SUSY}}$ from $1~\mathrm{TeV}$ to $10~\mathrm{TeV}$, however, only minor changes ($\lesssim 5\%$) are found on the mixing parameters.

In Fig.~\ref{fg:RGE_t3050}, the allowed regions of three neutrino mixing angles $\{\theta^{}_{12}, \theta^{}_{13}, \theta^{}_{23}\}$, two leptonic CP-violating phases $\{\delta, \sigma\}$ and two nonzero neutrino masses $\{m^{}_1, m^{}_2\}$ are shown in the MSSM with $\tan \beta = 30$ and $\tan \beta = 50$. The allowed parameter space at the low-energy scale is denoted by black points, and one can observe that $\delta$ and $\sigma$ are restricted to a small area around $\delta = 90^\circ$ and $\sigma = 10^\circ$. This observation can be easily understood with the help of Eq.~(\ref{eq: approx}), which indicates that the deviations of $(\delta, \sigma)$ from $(90^\circ, 0^\circ)$ are measured by the neutrino mass-squared difference $\Delta m^2_{21} = (1 - \zeta^2)m^2_2$ and the small but nonzero mixing angle $\theta^{}_{13}$. At the high-energy seesaw scale $\Lambda^{}_{\rm SS} = 10^{13}~{\rm GeV}$, the parameter space in the MSSM with $\tan \beta = 30$ and $\tan \beta = 50$ has been represented by dark- and light-gray points, respectively. One can see that the RG running effects on $\theta^{}_{13}$ and $\theta^{}_{23}$ are insignificant, whereas the running effects on $\theta^{}_{12}$, $\delta$ and $\sigma$ are indeed remarkable. Therefore, it is necessary to include the running effects on those parameters when we consider the generation of baryon number asymmetry in our Universe, which takes place at a superhigh-energy scale.

\begin{figure}[!t]
\begin{center}
\subfigure{%
\includegraphics[width=0.4\textwidth]{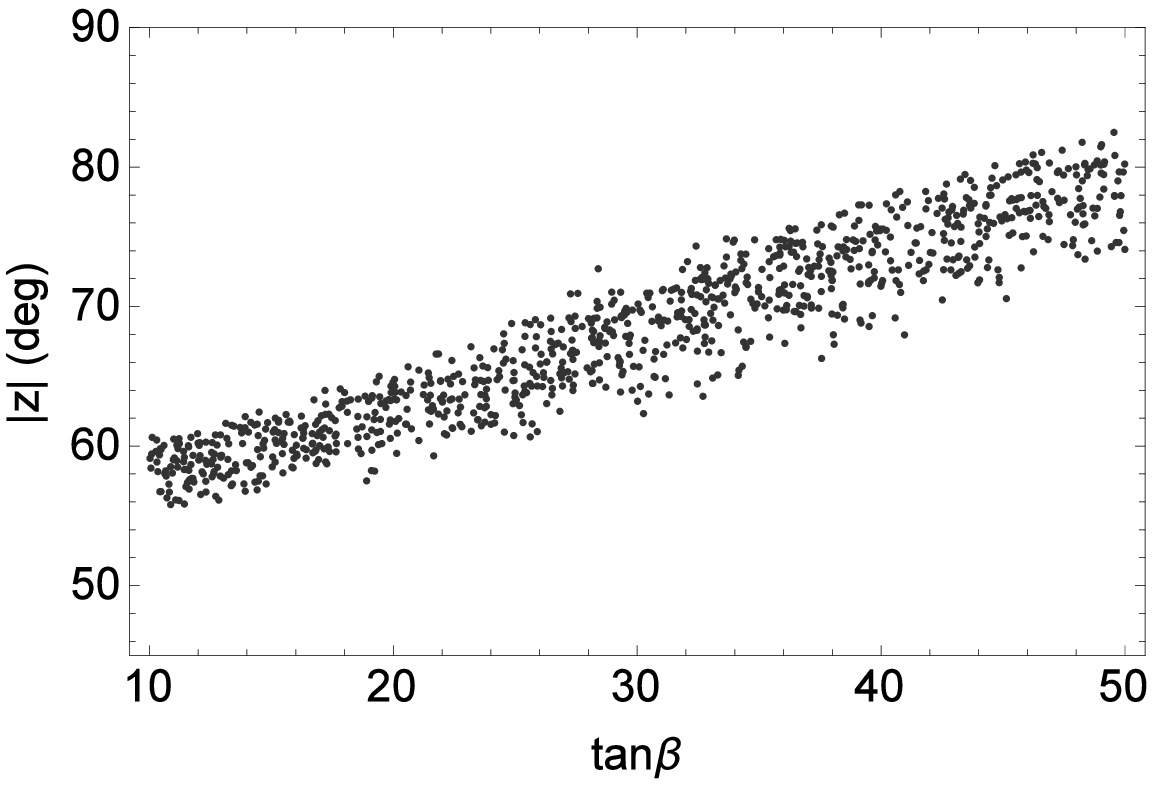}        }%
\subfigure{%
\hspace{0.8cm}
\includegraphics[width=0.4\textwidth]{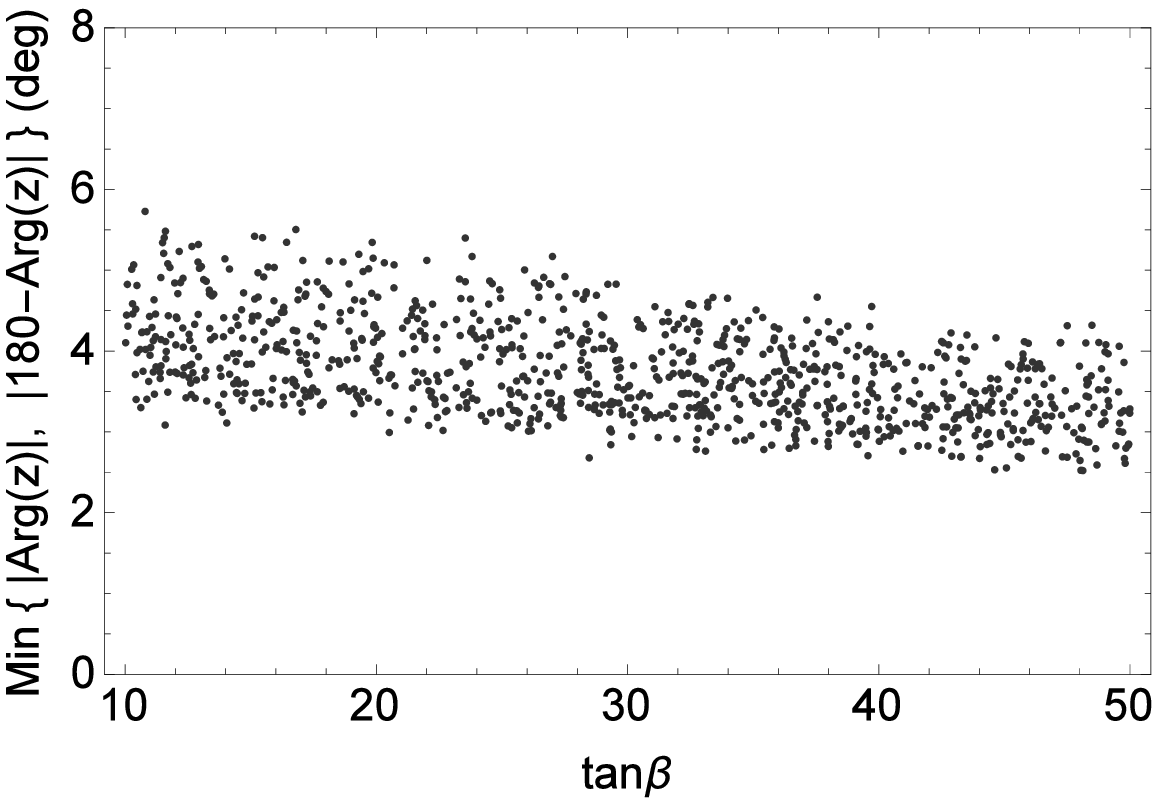}        }
\end{center}
\vspace{-0.3cm}
\caption{The absolute value $|z|$ and the phase $\arg{z}$ of the complex parameter $z$ are given in units of degrees in the left and right panels, respectively. For a given value of $\tan \beta$, $|z|$ and $\arg{z}$ are calculated by varying the low-energy parameters in their $3\sigma$ ranges and the high-energy scales from $10^8~{\rm GeV}$ to $10^{13}~{\rm GeV}$. Note that $z$ is almost real in all cases as indicated in the right panel.}
\label{fg: zpara}
\end{figure}

From Eq.~(\ref{eq: tanz}), we can figure out the real and imaginary parts of $z$ in terms of neutrino mixing parameters. More explicitly, we have
\begin{eqnarray}
{\rm Re} z &\approx& - \frac{1}{2} \left[\arctan \left(\frac{\sin \sigma + \cot \theta^{}_{12}}{\cos \sigma}\right) - \arctan \left(\frac{\sin \sigma - \cot \theta^{}_{12}}{\cos \sigma}\right)\right] \; , \nonumber \\
{\rm Im} z &\approx& - \frac{1}{4} \ln \left( \frac{1 - 2\sin \sigma \cot \theta^{}_{12} + \cot^2 \theta^{}_{12}}{1 + 2\sin \sigma \cot \theta^{}_{12} + \cot^2 \theta^{}_{12} }\right) \; ,
\label{eq: ReImz}
\end{eqnarray}
where $\zeta \approx 1$ is assumed. For a small $\tan \beta$, the RG running effects are negligible, so the mixing parameters can be identified with those extracted from oscillation experiments. In this case, one can expand Eq.~(\ref{eq: ReImz}) in terms of the Majorana CP-violating phase $\sigma$, which is constrained to be small. At the leading order, we get $|z| \approx \pi/2 - \theta^{}_{12}$ and $\arg{z} \approx \sigma \sin 2\theta^{}_{12} /(2\theta^{}_{12} - \pi)$. In the other extreme case, where the RG running is significant for a large $\tan \beta$, we can expand Eq.~(\ref{eq: ReImz}) in terms of $\theta^{}_{12}$ and obtain $|z| \approx \pi/2 - \cos \sigma \tan \theta^{}_{12}$ and $\arg{z} = 2 \sin \sigma \tan \theta^{}_{12}/\pi$. In both cases, $\arg{z}$ is found to be close to the real axis, i.e., around $5^\circ$. In general, both $\sigma$ and $\theta^{}_{12}$ are not small angles, and the above approximations are invalid.

However, one can compute the complex parameter $z$ by inputting the low-energy values of neutrino mixing parameters and solving the RG equations. The numerical results of $|z|$ and $\arg{z}$ are given in Fig.~\ref{fg: zpara}, where both small and large values of $\tan \beta$ are considered. Furthermore, the $3\sigma$ ranges of mixing parameters and a seesaw scale within $[10^8, 10^{13}]~{\rm GeV}$ are taken into account. One can see from the right panel of Fig.~\ref{fg: zpara} that a small phase of $z$ is obtained in all cases, implying the suppression of CP violation at the high-energy scale. The latter observation becomes clearer when we calculate the CP asymmetries in the decays of heavy Majorana neutrinos.

\section{Baryon Number Asymmetry}

One salient feature of the canonical seesaw model is to simultaneously explain tiny neutrino masses and the observed baryon number asymmetry in our Universe, which is usually measured by the baryon to photon density ratio~\cite{Ade:2013zuv}
\begin{equation}
\eta^0_{\rm B} \equiv \frac{n^{}_{\rm B}}{n^{}_\gamma} = (6.065 \pm 0.090) \times 10^{-10} \; ,
\label{eq: etaB0}
\end{equation}
where $n^{}_{\rm B}$ and $n^{}_\gamma$ stand for today's baryon and photon number density, respectively. In the very early Universe, when the reheating temperature after inflation is so high that heavy Majorana neutrinos $N^{}_i$ can be produced in thermal equilibrium. As the Universe cools down, the CP-violating decays of $N^{}_i$ will go out of thermal equilibrium if the decay rate becomes smaller than the expansion rate of the Universe. The CP asymmetries in the decays of $N^{}_i$ into leptons of different flavors are defined as~\cite{Buchmuller:2004nz,Buchmuller:2005eh,Davidson:2008bu}
\begin{equation}
\varepsilon^{}_{i\alpha} \equiv \frac{\Gamma(N^{}_i \to l^{}_\alpha H) - \Gamma(N^{}_i \to \overline{l}^{}_\alpha \overline{H})}{\Gamma(N^{}_i \to l^{}_\alpha H) + \Gamma(N^{}_i \to \overline{l}^{}_\alpha \overline{H})} \; ,
\label{eq: epsilon}
\end{equation}
where $\Gamma(N^{}_i \to l^{}_\alpha H)$ and $\Gamma(N^{}_i \to \overline{l}^{}_\alpha \overline{H})$ for $\alpha = e, \mu, \tau$ denote the decay rates of $N^{}_i$ into leptons $l^{}_\alpha$ and anti-leptons $\overline{l}^{}_\alpha$, respectively. It is the interference between the tree-level and one-loop decay amplitudes that gives rise to CP asymmetries, which receive both contributions from the one-loop self-energy and vertex corrections. More explicitly, we obtain
\begin{eqnarray}
\varepsilon_{i\alpha} = \frac{1}{8\pi (Y_{\nu}^\dagger Y_{\nu}^{})_{ii}} {\rm Im}  \sum_{k\neq i} (Y_{\nu}^*)_{\alpha i} (Y_{\nu}^{})_{\alpha k} \left[ (Y_{\nu}^\dagger Y_{\nu}^{})_{ik} f(x^{}_{ki}) + (Y_{\nu}^\dagger Y_{\nu}^{})^*_{ik} g(x^{}_{ki})\right] \; ,
\label{eq: CP}
\end{eqnarray}
where $x^{}_{ki} \equiv M^2_k/M^2_i$ and the loop functions are defined as follows
\begin{eqnarray}
f(x^{}_{ki}) &=& \sqrt{x_{ki}} \left[ \frac{1-x^{}_{ki}}{(1-x^{}_{ki})^2 + r^2_{ki}} + 1-(1+x_{ki}) \ln\frac{1+x_{ki}}{x_{ki}} \right] \; , \nonumber\\
g(x^{}_{ki}) &=& \frac{1 - x^{}_{ki}} {(1-x^{}_{ki})^2 + r^2_{ki}} \;.
\label{eq: fg}
\end{eqnarray}
If the mass spectrum of heavy Majorana neutrinos is strongly hierarchical, $r^{}_{ki}$ can be neglected in the denominators in Eq.~(\ref{eq: fg}). However, it serves as an important regulator to avoid any singularity in the limit of mass degeneracy $M^2_k = M^2_i$ or equivalently $x^{}_{ki} = 1$. In the resonant regime, the true form of $r^{}_{ki}$ is still controversial at present~\cite{Dev:2014laa}, and three distinct expressions have been derived: (i) $r^{}_{ki} = x^{}_{ki} \Gamma^{}_k/M^{}_k$ by a quantum field-theoretic approach~\cite{Pilaftsis:1997jf,Pilaftsis:2003gt}; (ii) $r^{}_{ki} = \Gamma^{}_i/M^{}_i - x^{}_{ki} \Gamma^{}_k/M^{}_k$ by a modified version~\cite{Buchmuller:1997yu,Anisimov:2005hr} of the approach introduced in Ref.~\cite{Pilaftsis:1997jf}; (iii) $r^{}_{ki} = \Gamma^{}_i/M^{}_i + x^{}_{ki} \Gamma^{}_k/M^{}_k$ by an effective Kadanoff-Baym approach with a specific quasi-particle ansatz~\cite{Garny:2011hg,Iso:2013lba}. As we numerically demonstrate in the FGY model, three different expressions of $r^{}_{ki}$ lead to the same result if a successful leptogenesis is realized.

The produced lepton-number asymmetries in the $N^{}_i$ decays will partly be washed out by the inverse decays and lepton-number-violating scattering, if these processes proceed efficiently. In order to describe the washout effects, we introduce the decay parameters $K^{}_i \equiv \Gamma^{}_i/H(M^{}_i)$, where $\Gamma^{}_i = (Y^\dagger_\nu Y^{}_\nu)^{}_{ii} M^{}_i/8\pi$ is the total decay width of $N^{}_i$ and $H(M^{}_i)$ is the Hubble parameter at temperature $T = M^{}_i$. In the radiation-dominated epoch, the Hubble parameter is given as a function of temperature $H(T) =1.66 \sqrt{g^*(T)} T^2/M^{}_{\rm pl}$, where $M^{}_{\rm pl} = 1.2\times 10^{19}~{\rm GeV}$ is the Planck mass and $g^*(T)$ is the number of relativistic degrees of freedom at $T$. The lepton number asymmetries will be converted into the baryon number asymmetry through the $(B+L)$-violating and $(B-L)$-conserving sphaleron processes~\cite{Manton:1983nd,Klinkhamer:1984di}, which are in thermal equilibrium between $T = 200~{\rm GeV}$ and $10^{12}~{\rm GeV}$. The final baryon number asymmetry is then given by~\cite{Buchmuller:2004nz}
\begin{equation}
\eta^{}_{\rm B} \approx -0.96\times 10^{-2} \sum_i \sum_\alpha \varepsilon^{}_{i \alpha} \kappa^{}_{i\alpha}
\label{eq: etaB}
\end{equation}
where the efficiency factors $\kappa^{}_{i\alpha}$ can be determined by solving the Boltzmann equations of heavy Majorana neutrino and lepton number densities. Roughly speaking, they are governed by the flavor-dependent decay parameters $K^{}_{i\alpha} \equiv P^{}_{i\alpha} K^{}_i$, where $P^{}_{i\alpha} = |(Y^{}_\nu)^{}_{\alpha i}|^2/(Y^\dagger_\nu Y^{}_\nu)^{}_{ii}$ stands for the projection probability of the final lepton state in $N^{}_i$ decays onto a specific lepton-flavor state.

So far, we have focused on leptogenesis in the SM. In the MSSM, the CP asymmetries in the decays of both $N^{}_i$ and its superpartner are twice larger, since the number of particles running in the loops are doubled. However, in the strong washout regime, the inverse decay rates are also doubly efficient, reducing the lepton asymmetries by a factor of two. In addition, the particle content is twice much in the MSSM, so we have the number of relativistic degrees of freedom $g^* = 228.75$ in the MSSM, while $g^* = 106.75$ in the SM. Altogether, the baryon number asymmetry in either strong or weak washout regime in the supersymmetric case is not much changed with respect to the non-supersymmetric case~\cite{Davidson:2008bu}.

In the vanilla scenario of leptogenesis, the mass spectrum of heavy Majorana neutrinos is taken to be hierarchical, and only the lightest Majorana neutrino $N^{}_1$ and the one-flavor approximation are considered. This is actually done for the FGY model in the previous papers~\cite{Frampton:2002qc,Guo:2003cc,Mei:2003gn,Harigaya:2012bw}, where a narrow mass range of the lightest heavy Majorana neutrino $M^{}_1 \sim 5\times 10^{13}~{\rm GeV}$ has been found in the IO case. In the following, we calculate the baryon asymmetry via a flavor-dependent leptogenesis by taking into account the lepton flavor effects and non-hierarchical mass spectra of heavy Majorana neutrinos.

\subsection{Lepton Flavor Effects}

The interaction rates associated with charged-lepton Yukawa couplings become larger than the expansion rate of the Universe at different temperatures, and thus affect the washout effects on lepton number asymmetries~\cite{Barbieri:1999ma, Endoh:2003mz, Abada:2006fw, Nardi:2006fx,Abada:2006ea}. For $M^{}_i \gtrsim 10^{12}~{\rm GeV}$, the leptogenesis mechanism works at the temperature $T \sim M^{}_i$, where all the charged-lepton Yukawa interactions are negligible compared to the expansion rate. Therefore, the lepton state produced in the decays also participates in the inverse decays and lepton-number-violating scattering. In this case, it is valid to treat leptons as a single flavor in both generation and washout of lepton number asymmetries. The relevant quantities are just the total CP asymmetry $\varepsilon^{}_i = \sum_\alpha \varepsilon^{}_{i\alpha}$ and the efficiency factor $\kappa^{}_i$, which is determined by the decay parameter $K^{}_i$. For $10^{12}~{\rm GeV} \gtrsim M^{}_i \gtrsim 10^9~{\rm GeV}$, the $\tau$ charged-lepton Yukawa interaction is in thermal equilibrium and able to single out the $\tau$ lepton flavor in the thermal bath. Therefore, one has to deal with two lepton flavors, namely the $\tau$ flavor and a combination of $e$ and $\mu$ flavors. The relevant parameters are the CP asymmetries $\varepsilon^{}_{i\tau}$ and $\varepsilon^{}_{i2} \equiv \varepsilon^{}_{ie} + \varepsilon^{}_{i\mu}$, and the efficiency factors $\kappa^{}_{i\tau}$ and $\kappa^{}_{i2}$, which are calculable by using $K^{}_{i\tau}$ and $K^{}_{i2} \equiv K^{}_{ie} + K^{}_{i\mu}$. For $M^{}_1 \lesssim 10^9~{\rm GeV}$, both $\tau$ and $\mu$ charged-lepton Yukawa interactions are efficient enough to recognize $\tau$ and $\mu$ flavors in the system, implying that a three-flavor treatment is necessary.

\begin{figure}
\centering
\includegraphics[scale=0.7]{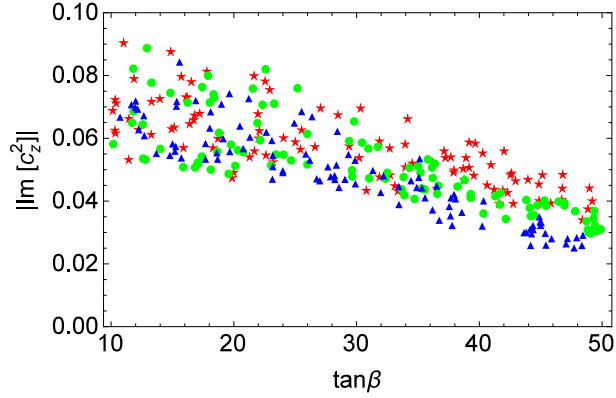}
\caption{Numerical results of ${\rm Im}[c^2_z]$ are calculated by solving the RG equations with the low-energy parameters in their $3\sigma$ ranges as inputs. The high-energy scale has been chosen to be $10^8~{\rm GeV}$ (red stars), $10^{10}~{\rm GeV}$ (green dots), and $10^{13}~{\rm GeV}$ (blue triangles) for a given $\tan \beta$ in the MSSM.}
\label{fg: ImZ}
\end{figure}

First, we compute the CP asymmetries in the FGY model. Since the Dirac neutrino Yukawa coupling matrix is given in Eq.~(\ref{eq: CI}), it is straightforward to figure out $\varepsilon^{}_{i\alpha}$ in Eq.~(\ref{eq: CP}). In the hierarchical limit of $M^{}_1 \ll M^{}_2$, we need to just focus on $\varepsilon^{}_{1\alpha}$ and assume that the lepton asymmetries generated from the decays of $N^{}_2$ have been washed out by the $N^{}_1$-related lepton-number-violating processes. For ${\bf Pattern ~B}^{}_1$ with $(Y^{}_\nu)^{}_{e1} = (Y^{}_\nu)^{}_{\mu 2} = 0$, we obtain $\varepsilon^{}_{1e} = \varepsilon^{}_{1\mu} = 0$, and
\begin{equation}
\varepsilon^{}_{1\tau} = \varepsilon^{}_1 \approx - \frac{3}{16\pi} \frac{M^{}_1}{v^2} \frac{\Delta m^2_{21} {\rm Im}[c^2_z]}{m^{}_1 |c^{}_z|^2 + m^{}_2 |s^{}_z|^2} \; ,
\label{eq: epsilon1}
\end{equation}
where the second equality has also been found in Ref.~\cite{Harigaya:2012bw}. The CP asymmetry is suppressed by the tiny neutrino mass-squared difference $\Delta m^2_{21} \approx 7.5\times 10^{-5}~{\rm eV}^2$. Furthermore, as we have shown in the previous section, the complex parameter $z$ is very close to the real axis, implying that $|{\rm Im}[c^2_z]| \approx |z| \sin(2|z|) \arg(z)$ should also be small. The numerical values of $|{\rm Im}[c^2_z]|$ have been presented in Fig.~\ref{fg: ImZ} for a wide range of model parameters, where one can observe that $|{\rm Im}[c^2_z]|$ is actually small and varies between $0.03$ and $0.09$. In the present work, we shall concentrate on ${\bf Pattern~B}^{}_1$, but one can calculate the CP asymmetries for the other three viable patterns in a similar way. The important results for all four viable patterns have been summarized in Table 2.

Second, instead of solving the complete set of Boltzmann equations, we apply the analytical formulas obtained in Ref.~\cite{fl} to estimate the efficiency factors. If the initial thermal abundance of heavy Majorana neutrinos is assumed, the efficiency factor is approximately given by~\cite{fl}
\begin{equation}
\kappa^{}_{i\alpha} \approx \frac{2}{K^{}_{i\alpha} z^{}_{\rm B}(K^{}_{i\alpha})} \left[ 1 - \exp\left( - \frac{K^{}_{i\alpha} z^{}_{\rm B}(K^{}_{i\alpha})}{2}\right)\right] \; ,
\label{eq: effi}
\end{equation}
where $z^{}_{\rm B}(K^{}_{i\alpha}) = 2 + 4 K^{0.13}_{i\alpha} \exp(-2.5/K^{}_{i\alpha})$. Hence the efficiency factors are completely fixed by the decay parameters $K^{}_{i\alpha}$, which are in turn determined by the flavor structure of $Y^{}_\nu$. For ${\bf Pattern ~B}^{}_1$, we get the total decay parameter
\begin{equation}
K^{}_1 = \frac{M^2_1 v^2 (m^{}_1 |c^{}_z|^2 + m^{}_2 |s^{}_z|^2)}{8\pi H(M^{}_1)} \approx 50 \; ,
\label{eq: K1}
\end{equation}
where ${\rm Im} z \ll 1$ and $m^{}_2 \approx m^{}_1 \approx 0.05~{\rm eV}$ have been used in the last step. The projection probability is determined by
\begin{equation}
\frac{P^{}_{1\tau}}{1 - P^{}_{1\tau}} = \frac{|(Y^{}_\nu)^{}_{\tau 1}|^2}{|(Y^{}_\nu)^{}_{\mu 1}|^2} = \frac{|U^{}_{\tau 1} \sqrt{m_1} c^{}_z + U^{}_{\tau 2} \sqrt{m_2} s^{}_z|^2}{|U^{}_{\mu 1} \sqrt{m_1} c^{}_z + U^{}_{\mu 2} \sqrt{m_2} s^{}_z|^2} = \tan \theta^{}_{23} \; ,
\label{eq: P1t1}
\end{equation}
where the identity $\tan z = - U^{}_{e1} \sqrt{m_1}/(U^{}_{e 2} \sqrt{m_2})$ has been implemented to significantly simplify the result. Given $\theta^{}_{23} \approx 45^\circ$, we arrive at $P^{}_{1\tau} \approx 0.5$ and $K^{}_{1\tau} \approx 25$. For comparison, we can also figure out $P^{}_{1\tau}$ for ${\bf Pattern~B}^{}_4$ with $(Y^{}_\nu)^{}_{e2} = (Y^{}_\nu)^{}_{\mu 1} = 0$. With the constraint $\tan z = - U^{}_{\mu 1} \sqrt{m_1}/(U^{}_{\mu 2} \sqrt{m_2})$, we have
\begin{equation}
\frac{P^{}_{1\tau}}{1 - P^{}_{1\tau}} = \frac{|(Y^{}_\nu)^{}_{\tau 1}|^2}{|(Y^{}_\nu)^{}_{e 1}|^2} = \frac{|U^{}_{\tau 1} \sqrt{m_1} c^{}_z + U^{}_{\tau 2} \sqrt{m_2} s^{}_z|^2}{|U^{}_{e 1} \sqrt{m_1} c^{}_z + U^{}_{e 2} \sqrt{m_2} s^{}_z|^2} \approx \frac{\tan^2 \theta^{}_{13}}{\cos \theta^{}_{23}} \; ,
\label{eq: P1t4}
\end{equation}
and thus $P^{}_{1\tau} \approx 0.05$ and $K^{}_{1\tau} = 2.5$, which are one order of magnitude smaller than the result in the previous case. Since $P^{}_{1\tau}$ in Eq.~(\ref{eq: P1t1}) or Eq.~(\ref{eq: P1t4}) depends mainly on $\theta^{}_{23}$ and $\theta^{}_{13}$, its value should be quite stable against the RG running.

With both the CP asymmetries and decay parameters, we are ready to find out the efficiency factors, and then baryon number asymmetry. The numerical results are summarized as follows:
\begin{itemize}
\item If $M^{}_1 \gtrsim 10^{12}~{\rm GeV}$, we can treat leptons as a single flavor, and the relevant quantities are the CP asymmetry $\varepsilon^{}_1 \approx - 2\times 10^{-6}~ (M^{}_1/ 10^{13}~{\rm GeV})$, which is identical to $\varepsilon^{}_{1\tau}$ as shown in Eq.~(\ref{eq: epsilon1}), and the efficiency factor $\kappa^{}_1 \approx 5\times 10^{-3}$ by inserting $K^{}_1 = 50$ into the analytical formula in Eq.~(\ref{eq: effi}). Putting all together, we obtain the baryon number asymmetry
    \begin{equation}
    \eta^{}_{\rm B} \approx -0.96\times 10^{-2} \varepsilon^{}_1 \kappa^{}_1 = 1.0 \times 10^{-10} \left(\frac{M^{}_1}{10^{13}~{\rm GeV}}\right) \; ,
    \end{equation}
    which is in agreement with the result in Ref.~\cite{Harigaya:2012bw}. Therefore, heavy Majorana neutrinos should be as heavy as $6 \times 10^{13}~{\rm GeV}$ to generate the correct baryon number asymmetry. Note that $|{\rm Im}[c^2_z]| = 0.05$ has been assumed in the above calculation, but it is evident from Fig.~\ref{fg: ImZ} that the RG running effects on mixing parameters can enhance or reduce this value by a factor of two, depending on $\tan \beta$.

\item If $M^{}_1 < 10^{12}~{\rm GeV}$, the CP asymmetry is given by the same formula $\varepsilon^{}_{1\tau} \approx - 2 \times 10^{-7} ~ (M^{}_1/ 10^{12}~{\rm GeV})$, which will be at least one order of magnitude smaller compared to the previous case. Since the flavor structure of $Y^{}_\nu$ under consideration indicates $\varepsilon^{}_{1e} = \varepsilon^{}_{1\mu} = 0$, there is no contribution from other lepton flavors to the lepton number asymmetries. The washout of lepton number asymmetries is now determined by $K^{}_{1\tau} = P^{}_{1\tau} K^{}_1 = 25$, leading to an efficiency factor $\kappa^{}_{1\tau} \approx 0.01$. Although there is an enhancement by a factor of two, the mass of the heavy Majorana neutrino is too small to provide a large enough CP asymmetry. If we turn to the case of ${\bf Pattern}~{\bf B}^{}_4$, the CP asymmetry remains the same and the efficiency factor is $\kappa^{}_{1\tau} \approx 0.2$, so we have the final baryon number asymmetry
    \begin{equation}
    \eta^{}_{\rm B} \approx -0.96\times 10^{-2} \varepsilon^{}_{1\tau} \kappa^{}_{1\tau} = 3.8 \times 10^{-10} \left(\frac{M^{}_1}{10^{12}~{\rm GeV}}\right) \; ,
    \end{equation}
    which is on the right order of magnitude even for $M^{}_1 = 10^{12}~{\rm GeV}$. However, it is worthwhile to point out that $M^{}_1 = 10^{12}~{\rm GeV}$ is on the edge of two-flavor approximation, when the coherence of lepton state in $N^{}_1$ decays may be destroyed by the $\tau$ Yukawa interaction. In this case, the classical Boltzmann equations are not accurate enough to give the correct answer, and the fully quantum Boltzmann equations should be applied~\cite{Garny:2011hg,Iso:2013lba,Dev:2014laa}. Hence the flavor effects may open a possibility to realize a successful leptogenesis even for a smaller $M^{}_1$.
\end{itemize}
For even smaller masses $M^{}_1 \ll 10^{12}~{\rm GeV}$, the CP asymmetries are significantly suppressed. It is impossible to explain the observed baryon number asymmetry in the FGY model, although the flavor effects tend to protect lepton number asymmetry from washout.

\subsection{Beyond Hierarchical Limit}

The high mass scale of heavy Majorana neutrinos causes the so-called naturalness or fine-tuning problem for the light Higgs boson mass \cite{Vissani:1997ys,Abada:2007ux,Xing:2009in,Farina:2013mla,Clarke:2015gwa}, and the gravitino overproduction problem if the model is supersymmetrized~\cite{Giudice:2003jh}. In Ref.~\cite{Clarke:2015gwa}, a detailed analysis of the naturalness problem in the type-I seesaw model yields an upper bound on the heavy Majorana neutrino masses, namely, $M^{}_1 < 4\times 10^7~{\rm GeV}$ and $M^{}_2 < 7\times 10^7~{\rm GeV}$. These upper bounds have been derived by requiring that the radiative corrections induced by heavy Majorana neutrinos to the Higgs boson mass should be around the TeV scale. Obviously, this bound is in contradiction with the requirement of $M^{}_1 \sim 10^{13}~{\rm GeV}$ for explaining the baryon number asymmetry in the FGY model. Therefore, it is well motivated to go beyond the hierarchical limit and consider both mild mass hierarchy and a nearly-degenerate mass spectrum.

In the mild hierarchy case, we take $M^{}_2$ to be a few times $M^{}_1$. For the later convenience of quantifying the level of mass degeneracy, we introduce a dimensionless parameter
\begin{eqnarray}
\Delta \equiv \frac{M^{}_2 - M^{}_1}{M^{}_2} \; ,
\label{eq: massdeg}
\end{eqnarray}
which is zero in the limit of exact mass degeneracy $M^{}_1 = M^{}_2$ and approaches one for $M^{}_2 \gg M^{}_1$, which is the case discussed in the previous subsection.

Because of a mild hierarchy between $M^{}_1$ and $M^{}_2$, both $N^{}_1$ and $N^{}_2$ participate in the production and washout processes of lepton number asymmetries. The evolution of these asymmetries therefore involves solving the Boltzmann equations with both $N^{}_1$ and $N^{}_2$, and the previously used analytic formula for estimating the efficient factor is no longer applicable. To obtain a rough estimation of the baryon number asymmetry in this mild hierarchy case, we next consider a simplified set of Boltzmann equations, where only the inverse-decay processes are included in the washout term. First, the evolution equations of $N^{}_1$ and $N^{}_2$ number densities are~\cite{fl}
\begin{eqnarray}
\frac{{\rm d} n^{}_{N^{}_i}}{{\rm d} z} &=& - D^{}_i(n^{}_{N^{}_i} - n_{N^{}_i}^{\rm eq}),
\label{eq: dN}
\end{eqnarray}
where $z=M^{}_1/T$, and $n_{N_i}^{}$ is the number density for $N_i$ normalized by its density in ultra-relativistic thermal equilibrium (i.e., $T \gg M^{}_i$). Here $n_{N^{}_i}^{\rm eq} = z_i^2 {\cal K}_2 (z_i)/2$ with $z_i \equiv M^{}_i/T = z M_i/M_1$ is the density in thermal equilibrium, and ${\cal K}_2(z)$ is the modified Bessel function of the second kind. The decay factor $D_i$ is defined to be
\begin{eqnarray}
D^{}_i \equiv \frac{\Gamma^{}_i(z)}{H(z) z} = K^{}_i  z \frac{M_i^2}{M_1^2} \langle \frac{1}{\gamma^{}_i} \rangle,
\end{eqnarray}
where $K^{}_i$ has the same form as the previously defined total washout factor, and $\langle 1/\gamma^{}_i \rangle = {\cal K}^{}_1(z^{}_i)/ {\cal K}^{}_2(z^{}_i)$ is the thermally averaged dilation factor. Second, we also have the evolution equations for the lepton asymmetries, namely,
\begin{eqnarray}
\frac{{\rm d} n^{}_{\Delta^{}_\alpha}}{{\rm d} z} = - \sum_i \varepsilon^{}_{i \alpha} D^{}_i (n^{}_{N^{}_i} - n_{N^{}_i}^{\rm eq}) - n^{}_{\Delta^{}_\alpha} \sum_i P^{}_{i\alpha} W_i^{\text{ID}} \; ,
\label{eq: dL}
\end{eqnarray}
where $n^{}_{\Delta_\alpha}$ is the $B - L$ asymmetry density for the flavor $\alpha$, which has also been normalized by the density of $N^{}_i$ in the ultra-relativistic thermal equilibrium, and the total $B-L$ asymmetry density $n^{}_{B-L}$ is then given by $n^{}_{B-L}= \sum_\alpha n^{}_{\Delta_\alpha}$. In addition, $P_{i\alpha}$ is the projection probability defined previously, and the inverse-decay washout term $W_i^{\text{ID}}$ is as follows
\begin{eqnarray}
W_i^{\text{ID}}=\frac{1}{4}K^{}_i\frac{M_i}{M_1}\mathcal{K}_1(z_i)z_i^3.
\end{eqnarray}

\begin{figure}
\centering
\includegraphics[scale=0.7]{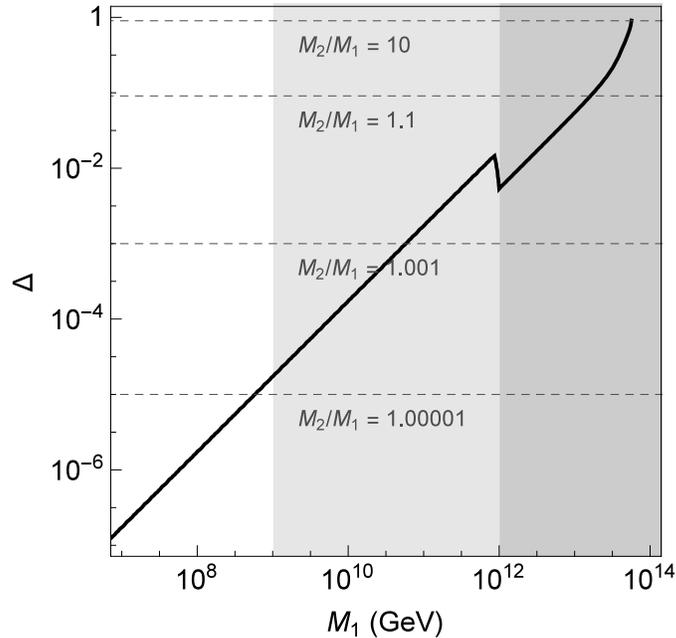}
\caption{Illustration for the dependence of baryon number asymmetry on the lightest heavy Majorana neutrino mass $M^{}_1$ and the mass degeneracy parameter $\Delta$. The black and solid curve corresponds to the allowed regions of model parameters, for which the observed baryon number asymmetry $\eta^{}_{\rm B} \approx 6.065 \times 10^{-10}$ can be naturally explained. The dashed lines indicate a few typical values of the mass ratio $M^{}_2/M^{}_1$.}
\label{fg: all}
\end{figure}

Given the above set of Boltzmann equations, we then solve them numerically. The initial conditions are obtained by setting the thermal abundance of $n^{}_{N^{}_i}$, and vanishing $B - L$ asymmetries. In Fig.~\ref{fg: all}, we present the allowed parameter space for $M^{}_1$ and $\Delta$ in the case of ${\bf Pattern}~{\bf B}^{}_1$. The black solid curve represents a contour of $\eta^{}_{\rm B} = 6.065\times 10^{-10}$, for which the observational uncertainty is so small that it will be hidden by the line width in the figure. The mass regions, which are represented by the shading areas, are characterized by the charged-lepton flavor effects.

In the highly degenerate case, we calculate $\eta_{\rm B}^{}$ in two ways: solving the simplified set of Boltzmann equations introduced Eqs.~(\ref{eq: dN}) and (\ref{eq: dL}), and applying the approximate analytical formulas. In \cite{flBHL}, it was argued that in the degenerate limit, the $N^{}_1$ and $N^{}_2$ washout contributions add up, resulting in
\begin{eqnarray}
\eta^{}_{\rm B} = -0.96\times 10^{-2} \sum_\alpha (\varepsilon^{}_{1\alpha}+\varepsilon^{}_{2\alpha})\kappa (K^{}_{1\alpha}+ K^{}_{2\alpha}),
\end{eqnarray}
where the efficiency factor $\kappa$ is still calculated via Eq.~(\ref{eq: effi}). The summation over $\alpha$ depends on the region of the lepton flavor effects. We focus on ${\bf Pattern}~{\bf B}^{}_1$ with $(Y^{}_\nu)_{e1} = (Y^{}_\nu)^{}_{\mu 2}=0$, and the other cases can be analyzed in a similar way.

In Fig.~\ref{fg: all}, we show the allowed parameter space for $M^{}_1$ and $\Delta$ for a variety of masses, in the single-flavor, two-flavor and three-flavor regions. We have demonstrated that the two approaches with simplified Boltzmann equations and approximate formulas lead to the same result. In the mild hierarchy case, we observe from Fig.~\ref{fg: all} that $M^{}_1$ still sits around $5\times 10^{13}~\text{GeV}$. This can be easily understood, as we know that $\varepsilon^{}_2$ is at most as large as $\varepsilon^{}_1$. To see this point clearly, we first calculate $\varepsilon^{}_1/\varepsilon^{}_2$ by using Eq.~(\ref{eq: epsilon}), and find it divergent when $\varepsilon^{}_2 =0$, corresponding to $M^{}_2/M^{}_1 \approx 2.36$. When $M^{}_2/M^{}_1 < 2.36$, $\varepsilon^{}_1$ and $\varepsilon^{}_2$ have the same sign, while the opposite situation happens when $M^{}_2/M^{}_1 > 2.36$. In addition, $|\varepsilon^{}_1| > |\varepsilon^{}_2|$ holds for all ratios of $M^{}_2/M^{}_1$. Therefore, including the contributions from $N^{}_2$ cannot significantly enhance the amount of CP asymmetry, and one then still needs to raise the mass scale of $M^{}_1$ so as to reach the required value of $\eta_{\rm B}^{}$. In the nearly-degenerate case, we see that a mass degeneracy at the level of $\Delta = 10^{-7}$ is required to meet the naturalness bound $M^{}_1 < 4\times 10^7~{\rm GeV}$ and account for the baryon number asymmetry via resonant leptogenesis~\cite{Pilaftsis:1997jf,Pilaftsis:2003gt,Xing:2006ms}. In our calculations, the formulas of CP asymmetries with different regulators $r^{}_{ik}$ lead to the same result in the FGY model. Although it seems unnatural to require such a high mass degeneracy, it can actually be achieved by implementing a flavor symmetry and its soft breaking at a superhigh-energy scale~\cite{Pilaftsis:2003gt}, or by the RG running effects~\cite{GonzalezFelipe:2003fi,Branco:2005ye}. As one can see, there is a kink around $M^{}_1 = 10^{12}~{\rm GeV}$. The reason is simply that we use different Boltzmann equations for the two cases of below and above $10^{12}~{\rm GeV}$. The kink should disappear if the fully quantum Boltzmann equations with coherent flavor effects are used~\cite{Dev:2014laa}. The curve is continuous around $M^{}_1 = 10^9~{\rm GeV}$, since the flavor structure enforces only one nonzero CP asymmetry $\varepsilon^{}_{1\tau}$.

\section{Summary}

In light of the latest neutrino oscillation data, we have performed a further study of the FGY model, in which only two right-handed heavy Majorana neutrinos are introduced and two texture zeros appear in the Dirac neutrino Yukawa coupling matrix, by taking into account the RG running of neutrino mixing parameters and flavor effects in leptogenesis. Such an investigation is well motivated in two aspects.

First, the FGY model is very interesting and predictive, and can be readily confronted with the latest neutrino oscillation data. Since the lightest neutrino is massless, the neutrino mass spectrum is fixed by the neutrino mass-squared differences, which are precisely measured in neutrino oscillation experiments. There are one Dirac and one Majorana CP-violating phases, which are actually determined by neutrino mixing angles and masses. The neutrino mass ordering is inverted, implying that the effective neutrino mass $m^{}_{\beta \beta} = 50~{\rm meV}$ is well within the reach of next-generation neutrinoless double-beta decay experiments.

Second, either the renormalization-group running effects of neutrino mixing parameters or the lepton flavor effects in leptogenesis has been ignored in the previous studies. Moreover, in order to stabilize the Higgs boson mass, the lightest heavy Majorana neutrino mass should be light enough $M^{}_1 < 4\times 10^7~{\rm GeV}$, which contradicts with the requirement $M^{}_1 \sim 10^{13}~{\rm GeV}$ for a successful leptogenesis. It is interesting to revisit this economical model by considering RG running effects, lepton flavor effects in leptogenesis and a non-hierarchical mass spectrum of heavy Majorana neutrinos.

In this work, taking account of the RG running effects on neutrino mixing parameters, we have consolidated the conclusions reached in Ref.~\cite{Harigaya:2012bw} and demonstrated that only four patterns ${\bf B}^{}_1$, ${\bf B}^{}_2$, ${\bf B}^{}_4$, and ${\bf B}^{}_5$ in Eq.~(\ref{eq: CaseB}) in the IO case are allowed by current neutrino oscillation data. This generalization is important for the MSSM with a large value of $\tan\beta$, where the RG running effects are significant. It has been found that the determination of neutrino mass ordering and the observation of neutrinoless double-beta decays will provide critical evidences to verify or disprove these four patterns. Furthermore, the octant of $\theta^{}_{23}$ and the CP-violating phase $\delta$ will be measured in future long-baseline neutrino oscillation experiments, and then can be used to further distinguish between ${\bf B}^{}_1$ (or ${\bf B}^{}_4$) and ${\bf B}^{}_2$ (or ${\bf B}^{}_5$). If the baryon number asymmetry is interpreted via leptogenesis mechanism, the relative sign of low-energy CP violation (i.e., the Jarlskog invariant ${\cal J} \propto \sin \delta$) to the high-energy CP violation (i.e., the CP asymmetry $\varepsilon^{}_1$ in $N^{}_1$ decays) serves as a discriminator for ${\bf B}^{}_1$ (${\bf B}^{}_2$) and ${\bf B}^{}_4$ (${\bf B}^{}_5$). The most important formulas for four viable patterns are collected in Table 2. If the naturalness criterion is applied to the FGY model, only the nearly-degenerate mass spectrum of heavy Majorana neutrinos with a mass degeneracy of $\Delta \sim 10^{-7}$ is allowed, and resonant leptogenesis becomes responsible for the baryon number asymmetry.

The FGY model actually exemplifies the idea of Occam's razor, which cuts away unnecessary free parameters and renders the model to be most economical and testable. If one of four viable patterns of the flavor structure is singled out by future neutrino experiments, we should go further to identify the underlying symmetries and explore the true dynamics for neutrino masses and lepton flavor mixing.

\vspace{0.5cm}

{\noindent \bf Acknowledgements}

\vspace{0.3cm}

This work was supported in part by the Innovation Program of the Institute of High Energy Physics under Grant No. Y4515570U1, and by the CAS Center for
Excellence in Particle Physics (CCEPP).

\newpage

\begin{landscape}

\begin{table}[t]
\begin{center}
\vspace{-0.25cm} \caption{Collection of important formulas for the four viable patterns.} \vspace{0.5cm}
{
\renewcommand{\arraystretch}{3.0}
\small
\begin{tabular}{c | c c c c}
\hline
\hline
& ${\bf Pattern}~{\bf B}^{}_1$ & ${\bf Pattern}~{\bf B}^{}_2$ & ${\bf Pattern}~{\bf B}^{}_4$ & ${\bf Pattern}~{\bf B}^{}_5$ \\
& ${\renewcommand{\arraystretch}{1.5} \left(\begin{matrix} {\bf 0}& \times \cr \times & {\bf 0} \cr \times & \times \end{matrix}\right)}$ & ${\renewcommand{\arraystretch}{1.5} \left(\begin{matrix} {\bf 0}& \times \cr \times & \times \cr \times & {\bf 0} \end{matrix}\right)}$ & ${\renewcommand{\arraystretch}{1.5} \left(\begin{matrix} \times & {\bf 0} \cr {\bf 0} & \times \cr \times & \times \end{matrix}\right)}$ & ${\renewcommand{\arraystretch}{1.5} \left(\begin{matrix} \times & {\bf 0} \cr \times & \times \cr {\bf 0} & \times \end{matrix}\right)}$ \\
\hline
$\cos\delta$ & $\displaystyle \frac{\sin 2\theta^{}_{12}(1 - \zeta^2)}{4 \tan \theta^{}_{23} \sin \theta^{}_{13}}  - \frac{\tan \theta^{}_{23}\sin \theta^{}_{13}}{\tan 2\theta^{}_{12}} $  & $\displaystyle  \frac{\cot \theta^{}_{23}\sin \theta^{}_{13}}{\tan 2\theta^{}_{12}} - \frac{\sin 2\theta^{}_{12}(1 - \zeta^2)}{4 \cot \theta^{}_{23} \sin \theta^{}_{13}}$  & $\displaystyle \frac{\sin 2\theta^{}_{12}(1 - \zeta^2)}{4 \tan \theta^{}_{23} \sin \theta^{}_{13}}  - \frac{\tan \theta^{}_{23}\sin \theta^{}_{13}}{\tan 2\theta^{}_{12}}$ &  $ \displaystyle \frac{\cot \theta^{}_{23}\sin \theta^{}_{13}}{\tan 2\theta^{}_{12}} - \frac{\sin 2\theta^{}_{12}(1 - \zeta^2)}{4 \cot \theta^{}_{23} \sin \theta^{}_{13}}$ \\
\hline
$\cos 2\sigma$ & $\displaystyle 1 - \frac{\tan^2 \theta^{}_{23} \sin^2 \theta^{}_{13}}{2 \sin^2 \theta^{}_{12} \cos^2 \theta^{}_{12}} $  & $\displaystyle 1 - \frac{\cot^2 \theta^{}_{23} \sin^2 \theta^{}_{13}}{2 \sin^2 \theta^{}_{12} \cos^2 \theta^{}_{12}}$  & $\displaystyle 1 - \frac{\tan^2 \theta^{}_{23} \sin^2 \theta^{}_{13}}{2 \sin^2 \theta^{}_{12} \cos^2 \theta^{}_{12}}$ & $\displaystyle 1 - \frac{\cot^2 \theta^{}_{23} \sin^2 \theta^{}_{13}}{2 \sin^2 \theta^{}_{12} \cos^2 \theta^{}_{12}}$\\
\hline
$\tan z$ & $\displaystyle - \frac{\sqrt{\zeta}}{\tan \theta^{}_{12}} e^{-{\rm i}\sigma}$  & $\displaystyle - \frac{\sqrt{\zeta}}{\tan \theta^{}_{12}} e^{-{\rm i}\sigma}$  & $ \displaystyle \frac{\tan \theta^{}_{12}}{\sqrt{\zeta}} e^{{\rm i}\sigma}$ & $\displaystyle \frac{\tan \theta^{}_{12}}{\sqrt{\zeta}} e^{{\rm i}\sigma}$\\
\hline
$P^{}_{1\alpha}$ & $\displaystyle P^{}_{1\tau} = \frac{\tan \theta_{23}^{}}{1 + \tan \theta_{23}^{}}$  & $\displaystyle P^{}_{1\mu} = \frac{\cot \theta_{23}^{}}{1 + \cot \theta_{23}^{}}$  & $\displaystyle P^{}_{1\tau} = \frac{\tan^2 \theta_{13}^{}}{\cos\theta_{23} + \tan^2 \theta_{13}^{}}$ & $\displaystyle P^{}_{1\mu} = \frac{\tan^2 \theta_{13}^{}}{\sin\theta_{23} + \tan^2 \theta_{13}^{}}$ \\
\hline
${\rm Im}[c^2_z]$ & $\displaystyle \frac{1}{2} \sin 2\theta^{}_{12} \tan \theta_{23}^{} \sin \theta_{13}^{} \sin\delta$  & $\displaystyle -\frac{1}{2} \sin 2\theta^{}_{12} \cot \theta_{23}^{} \sin \theta_{13}^{} \sin\delta$  & $\displaystyle - \frac{1}{2} \sin 2\theta^{}_{12} \tan \theta_{23}^{} \sin \theta_{13}^{} \sin\delta$ & $\displaystyle \frac{1}{2} \sin 2\theta^{}_{12} \cot \theta_{23}^{} \sin \theta_{13}^{} \sin\delta$ \\
\hline
$\displaystyle \mathrm{sgn}\left(\frac{\eta_{\rm B}^{}}{\sin\delta}\right)$ & $+$  & $-$  & $-$ & $+$ \\
\hline
\hline
\end{tabular}
}
\end{center}

\end{table}

\end{landscape}

\end{document}